\documentclass[a4paper]{llncs}

\usepackage{longtable}
\usepackage{proof}
\usepackage[table]{xcolor}
\usepackage{stmaryrd}
\usepackage{mathtools}
\usepackage{setspace}
\usepackage{cite}
\usepackage{color}
\usepackage{amssymb}
\usepackage{float}
\usepackage{enumitem}

\DeclareMathAlphabet\mathbfcal{OMS}{cmsy}{b}{n}
\let\mathcal\undefined \DeclareMathAlphabet{\mathcal}{OMS}{cmsy}{m}{n}
\newcommand{\flqs}{\ensuremath{\mathsf{4LQS}}}
\newcommand{\flqsr}{\ensuremath{\mathsf{4LQS^R}}}
\newcommand{\dlss}{\mathcal{DL}\langle \mathsf{4LQS^R}\rangle(\D)}
\newcommand{\dlssx}{\mathcal{DL}\langle \mathsf{4LQS^{R,\!\times}}\rangle(\D)}
\newcommand{\shdlssx}{\mathcal{DL}_{\D}^{4,\!\times}}
\newcommand{\shdlss}{\mathcal{DL}_{\D}^{4}}
\newcommand{\D}{\mathbf{D}}
\newcommand{\sroiqd}{\mathcal{SROIQ}(\D)}

\newcommand{\defAs}{\coloneqq}
\newcommand{\I}{\mathbf{I}}
\newcommand{\Ind}{\mathbf{Ind}}

\newcommand{\Ra}{\mathbf{R_A}}
\newcommand{\Rd}{\mathbf{R_D}}
\newcommand{\sym}{\mathsf{Sym}}
\newcommand{\asym}{\mathsf{Asym}}
\newcommand{\refl}{\mathsf{Ref}}
\newcommand{\irref}{\mathsf{Irref}}
\newcommand{\tra}{\mathsf{Tra}}
\newcommand{\fun}{\mathsf{Fun}}
\newcommand{\ck}{\mathsf{cpt}_\mathcal{KB}}
\newcommand{\ark}{\mathsf{arl}_\mathcal{KB}}
\newcommand{\crk}{\mathsf{crl}_\mathcal{KB}}
\newcommand{\ik}{\mathsf{ind}_\mathcal{KB}}

\newcommand{\bfk}{\mathsf{bf}_\mathcal{KB}^{\D}}
\newcommand{\vipcomment}[1]{}
\newcommand{\pow}{\mathcal{P}}

\newcommand{\T}{\mathcal{T}}
\newcommand{\seq}{\mathcal{S}^{\overline{\beta}}_i}

\newcommand{\seqnj}{\mathcal{S}^{\overline{\beta}}_j}
\newcommand{\ke}{KE-tableau}
\newcommand{\M}{\mathbfcal{M}}
\newcommand{\IM}{\I_\mathcal{M}}
\newcommand{\KB}{\mathcal{KB}}
\newcommand{\vari}{\mathtt{Var}_i}
\newcommand{\varz}{\mathtt{Var}_0}
\newcommand{\varu}{\mathtt{Var}_1}

\newcommand{\vart}{\mathtt{Var}_3}

\newcommand{\DT}{\mathcal{D}}

\newcommand{\dlmlsscart}{\ensuremath{\mathcal{DL\langle}\mathsf{MLSS}_{2,m}^{\times}\mathcal{\rangle}}}
\newcommand{\dlForallpizerotwo}{\ensuremath{\mathcal{DL\langle}\Forallpizerotwo\mathcal{\rangle}}}
\newcommand{\Forallpizerotwo}{\ensuremath{\mathbf{\forall_{0,2}^{\pi}}}}
			
\newtheorem{procedure}{Procedure}

\makeatletter
\def\@seccntformat#1{\@ifundefined{#1@cntformat}%
   {\csname the#1\endcsname\quad}  % default
   {\csname #1@cntformat\endcsname}% enable individual control
}
\let\oldappendix\appendix %% save current definition of \appendix
\renewcommand\appendix{%
    \oldappendix
    \newcommand{\section@cntformat}{\appendixname~\thesection\quad}
}
\makeatother

\title{Conjunctive Query Answering via a Fragment of Set Theory (Extended Version)}

\author{Domenico Cantone \and Marianna Nicolosi-Asmundo \and \\Daniele Francesco Santamaria}
\institute{
University of Catania, Dept. of Mathematics and Computer Science\\
~email:~\texttt{\{cantone,nicolosi,santamaria\}@dmi.unict.it}
}

\begin{document}
\maketitle

%\doublespace

\begin{abstract}
We address the problem of Conjunctive Query Answering (CQA) for the description logic $\dlssx$ ($\shdlssx$, for short) which extends the logic $\dlss$ with Boolean operations on concrete roles and with the product of concepts. 

The result is obtained by formalizing $\shdlssx$-knowledge bases and $\shdlssx$-conjunctive queries in terms of formulae of the four-level set-theoretic fragment 
$\flqsr$, which admits a restricted form of quantification on variables of the first three levels and on pair terms. We solve the CQA problem for $\shdlssx$ through a decision procedure for the satisfiability problem of $\flqsr$. We further define a \ke\space based procedure for the same problem, more suitable for implementation purposes, and analyze its computational complexity.  
%to saturate the so constructed knowledge base and resolve CQ entailment by means of the tableau provided. For the algorithm, we also provide 
%soundness and completeness proofs and analysis of complexity.
\end{abstract}

\section{Introduction}
In the last few years, results from Computable Set Theory %\cite{CaFeOm90,CaOmPo01,SchwCanOmo11} 
have been used as a means to represent and reason about description logics and rule languages for the semantic web. For instance, in \cite{CanLonNic2010, CanLonNic11, CanLon2014}, fragments of set theory with constructs related to \emph{multi-valued maps} have been studied and applied to the realm of knowledge representation.  In \cite{CanLonPis2010}, an expressive description logic, called $\dlmlsscart$, has been introduced and the consistency problem for $\dlmlsscart$-knowledge bases has been proved $\mathbf{NP}$-complete. The description logic $\dlmlsscart$ has been extended with additional constructs and SWRL rules in \cite{CanLonNic11}, proving that the decision problem for the resulting logic, called $\dlForallpizerotwo$, is still $\mathbf{NP}$-complete under suitable conditions. 
The description logic $\dlForallpizerotwo$ has been extended with some \emph{metamodelling} features in \cite{CanLon2014}. 
In \cite{CanLonNicSanRR2015}, the description logic $\dlss$ (more simply referred to as $\shdlss$) has been introduced. $\shdlss$ can be represented in the decidable four-level stratified fragment of set theory $\flqsr$ involving a restricted form of quantification over variables of the first three levels and pair terms (cf.\ \cite{CanNic2013}). The logic $\shdlss$ admits concept constructs such as full negation, union and intersection of concepts, concept domain and range, existential quantification and min cardinality on the left-hand side of inclusion axioms. It also supports role constructs such as role chains on the left hand side of inclusion axioms, union, intersection, and complement of abstract roles, and properties on roles such as transitivity, symmetry, reflexivity, and irreflexivity. It admits
datatypes, a simple form of concrete domains that are relevant in real world applications.  
%which turns out to be quite expressive if compared with $\sroiqd$, the description logic underpinning the Web Ontology Language OWL. 
The consistency problem for $\shdlss$-knowledge bases has been proved decidable in \cite{CanLonNicSanRR2015} by means of a reduction to the satisfiability problem for $\flqsr$, proved decidable in \cite{CanNic2013}. It has also been proved, under not very restrictive constraints, that the consistency problem for $\shdlss$-knowledge bases is \textbf{NP}-complete.
%$4LQS^R$ also provides a natural and tight integration of an expressive rules language.
Finally, we mention that the papers \cite{CanLonNic2010, CanLonNic11, CanLon2014,CanLonNicSanRR2015,CanLonPis2010} are concerned
with traditional research issues for description logics 
%in particular consistency of knowledge bases,  
mainly focused on the parts of a knowledge base representing conceptual information, namely the TBox and the RBox, where the principal reasoning services are subsumption and satisfiability. 

In this paper we exploit decidability results presented in \cite{CanNic2013,CanLonNicSanRR2015} to deal with reasoning services for knowledge bases involving ABoxes. The most basic service to query the instance data is \emph{instance retrieval}, i.e., the task of retrieving all individuals that instantiate a class $C$, and, dually, all named classes $C$ that an individual belongs to. In particular, a powerful way to query ABoxes is the \emph{Conjunctive Query Answering} task (CQA). CQA is relevant in the context of description logics and, in particular, for real world applications based on semantic web technologies, since it provides a mechanism allowing users and applications to interact with ontologies and data. The task of CQA has been studied for several well-known description logics (cf.\ \cite{Calvanese:1998:DQC:275487.275504, Calvanese2013335, calvanese2007answering, Ortiz:Calvanese:et-al:06a, GlHS07a,
GliHoLuSa-JAIR08,HorrSattTob-CADE-2000,j.websem63, HorrTess-aaai-2000,DBLP:conf/ijcai/HustadtMS05, Rosa07c, DBLP:conf/cade/Lutz08, Ortiz:2011:QAH:2283516.2283571}). 
In particular, we introduce the description logic $\dlssx$ ($\shdlssx$, for short), extending $\shdlss$ 
with Boolean operations on concrete roles and with the product of concepts. Then we define the CQA problem for $\shdlssx$ and prove its decidability via a reduction to the CQA problem for $\flqsr$, whose decidability follows from that of the satisfiability problem for $\flqsr$ (proved in \cite{CanNic2013}). Finally, we present a \ke\space based procedure that, given a $\shdlssx$-query $Q$ and a $\shdlssx$-knowledge base $\mathcal{KB}$ represented in set-theoretic terms, determines the answer set of $Q$ with respect to $\mathcal{KB}$, providing also some complexity results. The choice of the \ke\space system \cite{dagostino1999} is motivated by the fact that this variant of the tableau method allows one to construct trees whose distinct branches define mutually exclusive situations thus avoiding the proliferation of redundant branches, typical of semantic tableaux.

\section{Preliminaries}
\subsection{The set-theoretic fragment \flqsr} \label{4LQS}
%In order to define the fragment \flqsr, 
It is convenient to first introduce the syntax and semantics of a more general four-level quantified language, denoted $\flqs$.
Then %in Section \ref{4LQSR},
we provide some restrictions on quantified formulae of $\flqs$ that characterize \flqsr.
%
%
%present a subcollection of formulae of $4LQS$, called \flqsr, which is the fragment of our interest, characterized by a restricted form of quantification.
We recall that the satisfiability problem for \flqsr\ has been proved decidable in \cite{CanNic2013}.

%\subsubsection{Syntax of $\flqs$.}
$\flqs$ involves four collections, $\mathcal{V}_i$, of variables of sort $i$, for $i=0,1,2,3$. Variables of sort $i$, for $i=0,1,2,3$, will be denoted by $X^i,Y^i,Z^i,\ldots$ (in particular, variables of sort $0$ will also be denoted by $x, y, z, \ldots$).
In addition to variables, $\flqs$ involves also \emph{pair terms} of the form $\langle x,y \rangle$, with $ x,y \in \mathcal{V}_0$.

\smallskip

\noindent\emph{$\flqs$-quantifier-free atomic formulae} are classified as:
\begin{itemize}[topsep=0.1cm, itemsep=0.1cm]
\item[-] level 0:~~ $x=y$,~~ $x \in X^1$,~~ $\langle x,y \rangle = X^2$,~~ $\langle x,y \rangle \in X^3$;
\item[-] level 1:~~ $X^1=Y^1$,~~ $X^1 \in X^2$;
\item[-] level 2:~~ $X^2=Y^2$,~~ $X^2 \in X^3$.
\end{itemize}
%\end{itemize}
%(where, according to our conventions, $x,y \in \mathcal{V}_0$, $X^1,Y^1 \in \mathcal{V}_1$, $X^2,Y^2 \in \mathcal{V}_2$, and $X^3 \in \mathcal{V}_3$).

\smallskip

\noindent $\flqs$ \emph{purely universal formulae} are classified as:
%? topsep
%? partopsep 
%? parsep
%? itemsep
\begin{itemize}[topsep=0.1cm, itemsep=0.1cm]
\item[-] { level 1: $(\forall z_1)\ldots(\forall z_n) \varphi _0$, where $z_1,\ldots,z_n$  $\in \mathcal{V}_0$ and $\varphi _0$ is any propositional combination of quantifier-free atomic formulae of level 0;}
\item[-] { level 2: $(\forall Z^1_1)\ldots(\forall Z^1_m) \varphi _1$, where $Z^1_1,\ldots,Z^1_m $  $\in \mathcal{V}_1$ and $\varphi _1$ is any propositional combination of quantifier-free atomic formulae of levels 0 and 1, and of purely universal formulae of level 1;}
\item[-] {level 3: $(\forall Z^2_1)\ldots(\forall Z^2_p) \varphi _2$, where $Z^2_1,\ldots,Z^2_p $  $\in \mathcal{V}_2$ and $\varphi _2$ is any propositional combination of quantifier-free atomic formulae and of purely universal formulae of levels 1 and 2.}
\end{itemize}

\noindent
$\flqs$-formulae are all the propositional combinations of quantifier-free atomic formulae of levels 0, 1, 2 and of purely universal formulae of levels 1, 2, 3.

\medskip

Let $\varphi$ be a $\flqs$-formula. Without loss of generality, we can assume that $\varphi$ contains only $\neg$, $\wedge$, $\vee$ as propositional connectives. Further, let $S_{\varphi}$ be the syntax tree for a $\flqs$-formula $\varphi$,\footnote{The notion of syntax tree for $\flqs$-formulae is similar to the notion of syntax tree for formulae of first-order logic. A precise definition of the latter can be found in \cite{DeJo90}.} and let $\nu$ be a node of $S_{\varphi}$. We say that a $\flqs$-formula $\psi$ occurs within $\varphi$ at position $\nu$ if the subtree of $S_{\varphi}$ rooted at $\nu$
is identical to $S_{\psi}$. In this case we refer to $\nu$ as an occurrence of $\psi$ in $\varphi$ and to the path from the root
of $S_{\varphi}$ to $\nu$ as its occurrence path. An occurrence of $\psi$ within $\varphi$ is
\emph{positive} if its occurrence path deprived by its last node contains an even number of nodes labelled by a $\flqs$-formula of type $\neg \chi$. Otherwise, the occurrence is said to be \emph{negative}.

%nuova parte del 8 Giugno 
The variables $z_1,\ldots,z_n$ are said to occur \textit{quantified} in $(\forall z_1) \ldots (\forall z_n) \varphi_0$. Likewise, $Z^1_1,\ldots, Z^1_m$ and $Z^2_1, \ldots, Z^2_p$ occur quantified in $(\forall Z^1_1) \ldots (\forall Z^1_m) \varphi_1$ and in $(\forall Z^2_1) \ldots (\forall Z^2_p)  \varphi_2$, respectively.
A variable occurs \textit{free} in a $\flqs$-formula $\varphi$
%, and we say that $x$ is a free variable of $\varphi$, 
if it does not occur quantified in any subformula of $\varphi$. For $i = 0,1,2,3$, we denote with $\vari(\varphi)$ the collections of variables of level $i$ occurring free in $\varphi$.

A (level 0) substitution $\sigma \defAs \{ x_1/y_1, \ldots, x_n/y_n \}$ is the mapping $\varphi \mapsto \varphi\sigma$ such that, for any given $\flqs$-formula $\varphi$, $\varphi\sigma$ is the $\flqs$-formula obtained from $\varphi$ by replacing the free occurrences of the variables 
$x_1,\ldots, x_n$ in $\varphi$ with the variables $y_1, \ldots, y_n$, respectively. We say that a substitution $\sigma$ is free for $\varphi$ if the formulae $\varphi$ and $\varphi\sigma$ have exactly the same occurrences of quantified variables.  

%A (level 0) substitution $\sigma \defAs \{ x_1/y_1, \ldots, x_n/y_n \}$ is a mapping such that, given a $\flqs$-formula $\varphi$, $\varphi\sigma$ is the $\flqs$-formula obtained from $\varphi$ by replacing the \red{free occurrences of the} variables 
%$x_1,\ldots, x_n$ with the variables $y_1, \ldots, y_n$, respectively. We say that a substitution $\sigma$ is free for $\varphi$ if the formulae $\varphi$ and $\varphi\sigma$ have exactly the same occurrences of quantified variables.  

% % % % % % % % % % % % % % % % % % % % % % % % % %

A $\flqs$-\emph{interpretation} is a pair $\mathbfcal{M}=(D,M)$ where $D$ is a non-empty collection of objects (called \emph{domain} or \emph{universe} of $\mathbfcal{M}$) and $M$ is an assignment over the variables in $\mathcal{V}_i$, for $i=0,1,2,3$,  such that:\\[.1cm]
\centerline{$MX^{0} \in D, ~~~  MX^1 \in \pow(D), ~~~ MX^2 \in \pow(\pow(D)), ~~~ MX^3 \in \pow(\pow(\pow(D))),$}\\[.1cm]
where $ X^{i} \in \mathcal{V}_i$, for $i=0,1,2,3$, and $\pow(s)$ denotes the powerset of $s$.
%
%\smallskip
%{- $Mx \in D$, for each $ x \in \mathcal{V}_0$; $MX^1 \in \pow(D)$, for each $X^1 \in \mathcal{V}_1$;}
%
%%{- $MX^1 \in \pow(D)$, for each $X^1 \in \mathcal{V}_1$;}
%
%{- $ MX^2 \in \pow(\pow(D))$, for each $X^2 \in \mathcal{V}_2$; }
%
%{- $MX^3 \in \pow(\pow(\pow(D)))$, for each $X^3 \in \mathcal{V}_3$
%
%(we recall that $\pow(s)$ denotes the powerset of $s$).}

\smallskip
\noindent
Pair terms are interpreted \emph{\`a la} Kuratowski, and therefore we put \\[.1cm]
\centerline{$M \langle x,y \rangle \defAs \{ \{ Mx \},\{ Mx,My \} \}$.}
%\footnote{We recall that an \'a la Kuratowski ordered pair $(x,y)$ is defined as $(x,y)_K := \{\{x\}, \{x,y\}\}$.}
Next, let
\begin{itemize}[topsep=0.1cm, itemsep=0cm]
\item[-] $\mathbfcal{M}=(D,M)$ be a $\flqs$-interpretation,

\item[-] $x_1,\ldots,x_n \in \mathcal{V}_0$,~~ $X^1 _1, \ldots, X^1 _m \in \mathcal{V}_1$,~~ $X^2 _1, \ldots, X^2 _p \in \mathcal{V}_2$, and

\item[-] $u_1, \ldots, u_n \in D$,~~ $U^1 _1, \ldots, U^1 _m \in \pow(D)$,~~ $U^2 _1, \ldots, U^2 _p \in \pow(\pow(D))$.
\end{itemize}

%\smallskip
%- $\mathbfcal{M}=(D,M)$ be a $\flqs$-interpretation,
%
%- $x_1,....x_n \in \mathcal{V}_0$,~~ $X^1 _1, ... X^1 _m \in \mathcal{V}_1$,~~ $X^2 _1, ... X^2 _p \in \mathcal{V}_2$, and
%
%- $u_1, ... u_n \in D$,~~ $U^1 _1, ... U^1 _m \in \pow(D)$,~~ $U^2 _1, ... U^2 _p \in \pow(\pow(D))$.
%\renewcommand{\vec}[1]{\mathbf{#1}}
\smallskip
\noindent
By $\mathbfcal{M}[ \vec{x}  / \vec{u}, \vec{X}^1 / \vec{U}^1, \vec{X}^2 / \vec{U}^2 ]$,
%By  $\mathbfcal{M}[ x_1 / u _1, ..., x_n  / u_n, X^1 _1 / U^1 _1, ... X^1_m / U^1_m, X^2_1 / U^2_1, ... X^2_p / U^2_p  ] $, 
we denote the interpretation $\mathbfcal{M}'=(D,M')$ such that $M'x_i =u_i$ (for $i=1,\ldots,n$), $M'X^1_j = U^1 _j$ (for $j=1,\ldots,m$), $M'X^2_k = U^2_k$ (for $k=1,\ldots,p$), and which otherwise coincides with $M$ on all remaining variables. For a $\flqs$-interpretation $\mathbfcal{M} =(D,M)$ and a $\flqs$-formula $\varphi$, the satisfiability relationship $ \mathbfcal{M} \models \varphi$ is defined inductively over the structure of $\varphi$ as follows. Quantifier-free atomic formulae are evaluated in a standard way according to the usual meaning of the predicates `$\in$'
and `$=$', and purely universal formulae are evaluated as follows:
\begin{itemize}[topsep=0.2cm, itemsep=0.1cm]
\item[-] $\mathbfcal{M}  \models (\forall z_1) \ldots (\forall z_n) \varphi _0$ iff $\mathbfcal{M}   [ \vec{z} / \vec{u}] \models \varphi_0$, for all $\vec{u} \in D^{n}$; 

\item[-] $\mathbfcal{M}  \models (\forall Z^1_1) \ldots (\forall Z^1_m) \varphi _1$ iff $\mathbfcal{M}   [ \vec{Z}^{1} / \vec{U}^{1}] \models \varphi_1$, for all $\vec{U}^{1} \in \big(\pow(D)\big)^{m}  $;

\item[-] $\mathbfcal{M}  \models (\forall Z^2_1) \ldots (\forall Z^2_p) \varphi _2$ iff $\mathbfcal{M}   [ \vec{Z}^{2} / \vec{U}^{2}] \models \varphi_2$, for all $\vec{U}^{2} \in \big(\pow(\pow(D))\big)^{p}$.
\end{itemize}

Finally, compound formulae are interpreted according to the standard rules of propositional logic. If $\mathbfcal{M} \models \varphi$, then $\mathbfcal{M} $ is said to be a $\flqs$-model for $\varphi$. A $\flqs$-formula is said to be \emph{satisfiable} if it has a $\flqs$-model. A $\flqs$-formula is \emph{valid} if it is satisfied by all $\flqs$-interpretations. 
%Let $\varphi$ and $\psi$ be $\flqs$-formulae. 
%$\psi$ is said a \textit{logical consequence} of $\varphi$ (we write $\varphi \models \psi$), if $\M \models \varphi$ implies $\M \models \psi$, for every $\flqs$-interpretation $\M$.

We are now ready to present the fragment \flqsr\ of $\flqs$ of our interest. This is the collection of the formulae $\psi$ of $\flqs$ fulfilling the restrictions:
% Next we present the fragment \flqsr\ of $\flqs$ of our interest, namely the collection of the formulae $\psi$ of $\flqs$ fulfilling the restrictions:
\begin{enumerate}[label=\arabic*., topsep=0.1cm, itemsep=0.1cm]
\item for every purely universal formula $(\forall Z^1_1)\ldots(\forall Z^1_m) \varphi_1$ of level 2 occurring in $\psi$ and every purely universal formula $(\forall z_1)\ldots(\forall z_n) \varphi_0$ of level 1 occurring negatively in $\varphi_1$, $\varphi_0$ is a propositional combination of quantifier-free atomic formulae of level $0$ and the condition\\[.1cm]
\centerline{$\neg \varphi_0 \rightarrow \overset{n}{ \underset {i=1} \bigwedge} \; \overset {m} { \underset {j=1 }\bigwedge} z_i \in Z^1_j$}\\[.1cm]
%
%\[ {\small
%\neg \varphi_0 \rightarrow \overset{n}{ \underset {i=1} \bigwedge} \; \overset {m} { \underset {j=1 }\bigwedge} z_i \in Z^1_j }\]
is a valid $\flqs$-formula (in this case we say that $(\forall z_1)\ldots(\forall z_n) \varphi_0$ is \emph{linked to the variables} $Z^1_1,\ldots,Z^1_m$);

\item for every purely universal formula  $(\forall Z^2_1)\ldots(\forall Z^2_p) \varphi_2$  of level 3 in $\psi$:
\begin{itemize}[topsep=0.1cm, itemsep=0.cm]
\item[-] every purely universal formula of level 1 occurring negatively in $\varphi_2$ and not occurring in a purely universal formula of level 2 is only allowed to be of the form\\[.1cm]
\centerline{$(\forall z_1)\ldots(\forall z_n) \neg( \overset {n}{ \underset {i=1} \bigwedge} \; \overset {n} { \underset {j=1}\bigwedge} \langle z_i,z_j \rangle=Y^2_{ij}),$}\\[.1cm]
%\[
%(\forall z_1)\ldots(\forall z_n) \neg( \overset {n}{ \underset {i=1} \bigwedge} \; \overset {n} { \underset {j=1}\bigwedge} \langle z_i,z_j \rangle=Y^2_{ij}),\]
with $Y^2_{ij} \in \mathcal{V}^2$, for $i,j=1,\ldots,n$;

\item[-] purely universal formulae $(\forall Z^1_1)\ldots(\forall Z^1_m) \varphi_1$ of level 2 may occur only positively in $\varphi_2$.
\end{itemize}

\end{enumerate}

Restriction 1 has been introduced for technical reasons concerning the decidability of the satisfiability problem for the fragment, while 
%In fact it guarantees that satisfiability is preserved in a suitable finite submodel of $\psi$.  
restriction  2 allows one to define binary relations and several operations on them %while keeping simple, at the same time, the decision procedure 
(for space reasons details are not included here but can be found in \cite{CanNic2013}).

%We observe that 
The semantics of \flqsr\space plainly coincides with that of $\flqs$.

\subsection{The logic $\dlssx$}\label{dlssx}
The description logic $\dlssx$ (more simply referred to as $\shdlssx$) is the extension of the  logic $\dlss$ (for short $\shdlss$) presented in \cite{CanLonNicSanRR2015} in which Boolean operations on concrete roles and the product of concepts are admitted. Analogously to $\shdlss$, the logic $\shdlssx$
%that can be represented in the decidable four-level stratified fragment of set theory \flqsr. The logic $\shdlss$
supports concept constructs such as full negation, union and intersection of concepts, concept domain and range, existential quantification and min cardinality on the left-hand side of inclusion axioms, role constructs such as role chains on the left hand side of inclusion axioms, union, intersection, and complement of roles, and properties on roles such as transitivity, symmetry, reflexivity, and irreflexivity. 

As far as the construction of role inclusion axioms is concerned, $\shdlssx$ is more liberal than $\sroiqd\space$ (the logic underlying the most expressive Ontology Web Language 2 profile, OWL 2 DL \cite{owl2spec}), since the roles involved are not required to be subject to any ordering relationship, and the notion of simple role is not needed. 
$\shdlssx$ treats derived datatypes by admitting datatype terms constructed from data ranges by means of a finite number of applications of the Boolean operators. Basic and derived datatypes can be used inside inclusion axioms involving concrete roles.
%Such characteristics of $\shdlss$ make it suitable to express SWRL rules. The issue will be discussed in the next section.

Datatypes are defined according to \cite{Motik2008} as follows. Let $\D = (N_{D}, N_{C},N_{F},\cdot^{\D})$ be a \emph{datatype map}, where  $N_{D}$ is a finite set of datatypes, $N_{C}$ is a map assigning a set of constants $N_{C}(d)$ to each datatype $d \in N_{D}$, $N_{F}$ is a map assigning a set of facets $N_{F}(d)$ to each $d \in N_{D}$, and $\cdot^{\D}$ is a map assigning 

\begin{itemize} [itemsep=0.2cm]
\item [(i)] a datatype interpretation $d^{\D}$ to each datatype $d \in N_{D}$,
\item [(ii)] a facet interpretation $f^{\D} \subseteq d^{\D}$ to each facet $f \in N_{F}(d)$, and
\item [(iii)]  a data value $e_{d}^{\D} \in d^{\D}$ to every constant $e_{d} \in N_{C}(d)$.
\end{itemize}

We shall assume that the interpretations of the datatypes in $N_{D}$ are non-empty pairwise disjoint sets.

A \emph{facet expression} for a datatype $d \in N_{D}$ is a formula $\psi_d$ constructed from the elements of $N_{F}(d) \cup \{\top_{d},\bot_{d}\}$ by applying a finite number of times the connectives $\neg$, $\wedge$, and $\vee$. The function  $\cdot^{\D}$ is extended to facet expressions for $d  \in N_{D}$ by putting for $f, f_1,f_2 \in N_{F}(d)$

\begin{itemize}
\item [-] $\top_{d}^{\D} = d^{\D}$,
\item [-] $\bot_{d}^{\D} = \emptyset$,
\item [-] $(\neg f)^{\D} = d^{\D} \setminus f^{\D}$,
\item [-] $(f_1 \wedge f_2)^{\D} = f_1^{\D} \cap f_2^{\D}$,
\item [-] $(f_1 \vee f_2)^{\D} = f_1^{\D} \cup f_2^{\D}$.
\end{itemize}

 A \emph{data range} $dr$ for $\D$ is either a datatype $d \in N_{D}$, or a finite enumeration of datatype constants $\{e_{d_1},\ldots,e_{d_n}\}$, with $e_{d_i} \in N_{C}(d_i)$ and $d_i \in N_{D}$, or a facet expression $\psi_d$, for $d \in N_{D}$, or their complementation.

%\subsubsection{Syntax and Semantics of $\dlssx$.}
Let $\Ra$, $\Rd$, $\mathbf{C}$, $\Ind$ be denumerable pairwise disjoint sets of abstract role names, concrete role names, concept names, and individual names, respectively. We assume that the set of abstract role names $\Ra$ contains a name $U$ denoting the universal role. 
%
%The set of abstract roles is defined as $\Ra \cup \{ R^- \mid R \in \Ra \} \cup U$, where $U$ is the universal role and $R^-$ is the inverse role of $R$.

%%%%%%%%%%%%%%%%%%%%%%%%%%%%%%%%%%%%%%%%%%%%%%%%%%%%%%%%%%%%%%%%%%%%%%%%%%%%%%%%%%
 \vipcomment{An abstract role hierarchy $\mathsf{R}_{a}^{H}$ is a finite collection of RIAs.  A strict partial order $\prec$ on  $\Ra \cup \{ R^- \mid R \in \Ra \}$ is called \emph{a regular order} if $\prec$ satisfies, additionally, $S \prec R$ iff $S^- \prec R$, for all roles R and S.\footnote{We recall that a strict partial order $\prec$  on a set $A$ is an irreflexive and transitive relation on $A$.}}

\noindent
(a) $\shdlssx$-datatype, (b) $\shdlssx$-concept, (c) $\shdlssx$-abstract role, and (d) $\shdlssx$-concrete role terms are constructed according to the following syntax rules:
{
\begin{itemize}[topsep=0.4cm, itemsep=0.2cm]
\item[(a)] $t_1, t_2 \longrightarrow dr ~|~\neg t_1 ~|~t_1 \sqcap t_2 ~|~t_1 \sqcup t_2 ~|~\{e_{d}\}\, ,$

\item[(b)] $C_1, C_ 2 \longrightarrow A ~|~\top ~|~\bot ~|~\neg C_1 ~|~C_1 \sqcup C_2 ~|~C_1 \sqcap C_2 ~|~\{a\} ~|~\exists R.\mathit{Self}| \exists R.\{a\}| \exists P.\{e_{d}\}\, ,$

\item[(c)] $R_1, R_2 \longrightarrow S ~|~U ~|~R_1^{-} ~|~ \neg R_1 ~|~R_1 \sqcup R_2 ~|~R_1 \sqcap R_2 ~|~R_{C_1 |} ~|~R_{|C_1} ~|~R_{C_1 ~|~C_2} ~|~id(C) ~|~ $

$C_1 \times C_2    \, ,$

\item[(d)] $P_1,P_2 \longrightarrow T ~|~\neg P_1 ~|~ P_1 \sqcup P_2 ~|~ P_1 \sqcap P_2  ~|~P_{C_1 |} ~|~P_{|t_1} ~|~P_{C_1 | t_1}\, ,$
\end{itemize}
}
\noindent where $dr$ is a data range for $\D$, $t_1,t_2$ are data-type terms, $e_{d}$ is a constant in $N_{C}(d)$, $a$ is an individual name, $A$ is a concept name, $C_1, C_2$ are $\shdlssx$-concept terms, $S$ is an abstract role name,  $R, R_1,R_2$ are $\shdlssx$-abstract role terms, $T$ is a concrete role name, and $P,P_1,P_2$ are $\shdlssx$-concrete role terms. %We remark that data-type terms are introduced in order to represent derived data-types.

A $\shdlssx$-knowledge base is a triple ${\mathcal KB} = (\mathcal{R}, \mathcal{T}, \mathcal{A})$ such that $\mathcal{R}$ is a $\shdlssx$-$\mathit{RBox}$, $\mathcal{T}$ is a $\shdlssx$-$\mathit{TBox}$, and $\mathcal{A}$ a $\shdlssx$-$\mathit{ABox}$ (see next).

A $\shdlssx$-$\mathit{RBox}$ is a collection of statements of the following forms:\\[.4cm]
{
\centerline{
$R_1 \equiv R_2$, \hfill $R_1 \sqsubseteq R_2$, \hfill $R_1\ldots R_n \sqsubseteq R_{n+1}$, \hfill $\sym(R_1)$,\hfill $\asym(R_1)$,  } \\[.07cm]
\centerline{ \hfill $\refl(R_1)$, \hfill $\irref(R_1)$,\hfill $\mathsf{Dis}(R_1,R_2)$, \hfill $\tra(R_1)$, \hfill $\fun(R_1)$,  }\\[.07cm]
\centerline{\hfill $R_1 \equiv C_1 \times C_2$, \hfill $P_1 \equiv P_2$,  \hfill $P_1 \sqsubseteq P_2$, \hfill $\mathsf{Dis}(P_1,P_2)$, \hfill$\fun(P_1)$,}
}\\[.1cm]

%$R_1 \equiv R_2$, $R_1 \sqsubseteq R_2$, $R_1\ldots R_n \sqsubseteq R_{n+1}$, $\sym(R_1)$, $\asym(R_1)$, $\refl(R_1)$, $\irref(R_1)$, $\mathsf{Dis}(R_1,R_2)$,
%$\tra(R_1)$, $\fun(R_1)$, $R_1 \equiv C_1 \times C_2$, $P_1 \equiv P_2$, $P_1 \sqsubseteq P_2$, $\mathsf{Dis}(P_1,P_2)$, $\fun(P_1)$, 
where $R_1,R_2$ are $\shdlssx$-abstract role terms, $C_1, C_2$ are $\shdlssx$-abstract concept terms, and $P_1,P_2$ are $\shdlssx$-concrete role terms. Any expression of the type $w \sqsubseteq R$, where $w$ is a finite string of $\shdlssx$-abstract role terms and $R$ is an $\shdlssx$-abstract role term is called a \emph{role inclusion axiom (RIA)}. 

Next, a $\shdlssx$-$\mathit{TBox}$ is a set of statements of the types:\\[.3cm]
{
\centerline{$C_1 \equiv C_2$, \hfill$C_1 \sqsubseteq C_2$, \hfill$C_1 \sqsubseteq \forall R_1.C_2$, \hfill$\exists R_1.C_1 \sqsubseteq C_2$, \hfill$\geq_n\!\! R_1. C_1 \sqsubseteq C_2$, \hfill$C_1 \sqsubseteq {\leq_n\!\! R_1. C_2}$, }\\[.08cm]
\centerline{$t_1 \equiv t_2$, \hfill$t_1 \sqsubseteq t_2$, \hfill$C_1 \sqsubseteq \forall P_1.t_1$, \hfill$\exists P_1.t_1 \sqsubseteq C_1$, \hfill$\geq_n\!\! P_1. t_1 \sqsubseteq C_1$, \hfill$C_1 \sqsubseteq {\leq_n\!\! P_1. t_1}$,}
}\\[.3cm]
%\begin{itemize}[topsep=0.1cm, itemsep=0.cm]
%\item[-]  $C_1 \equiv C_2$,~~ $C_1 \sqsubseteq C_2$,~~ $C_1 \sqsubseteq \forall R_1.C_2$,~~ $\exists R_1.C_1 \sqsubseteq C_2$,~~ $\geq_n\!\! R_1. C_1 \sqsubseteq C_2$, \\$C_1 \sqsubseteq {\leq_n\!\! R_1. C_2}$,
%\item[-] $t_1 \equiv t_2$,~ $t_1 \sqsubseteq t_2$,~ $C_1 \sqsubseteq \forall P_1.t_1$,~ $\exists P_1.t_1 \sqsubseteq C_1$,~ $\geq_n\!\! P_1. t_1 \sqsubseteq C_1$,~ $C_1 \sqsubseteq {\leq_n\!\! P_1. t_1}$,
%\end{itemize}
%\marginpar{dire che t permette la costruzione di datatype misti}
where $C_1,C_2$ are $\shdlssx$-concept terms, $t_1,t_2$ datatype terms, $R_1$  a $\shdlssx$-abstract role term, and $P_1$ a $\shdlssx$-concrete role term.  Any statement $C \sqsubseteq D$, with  $C$ and $D$ $\shdlssx$-concept terms, is a 
\emph{general concept inclusion axiom (GCI)}.

Finally, a $\shdlssx$-$\mathit{ABox}$ is a set of \emph{individual assertions} of the forms:\\[0.3cm]
\centerline{$a : C_1$,~~ $(a,b) : R_1$, ~~
%$(a,b) : \neg R_1$, 
$a=b$, ~~ 
$e_{d} : t_1$,~~ 
$(a, e_{d}) : P_1$,}\\[0.3cm]
%$(a, e_{d}) : \neg P_1$, %$e_{d_1} = e_{d_2}$, $e_{d_1} \neq e_{d_2}$,
with $a,b$ individual names, $C_1$ a $\shdlssx$-concept term, $R_1$ a $\shdlssx$-abstract role term, $d$ a datatype, $e_{d}$ a constant in $N_{C}(d)$, $t_1$ a datatype term, and $P_1$ a $\shdlssx$-concrete role term.

The semantics of $\shdlssx$ is based on interpretations $\I= (\Delta^\I, \Delta_{\D}, \cdot^\I)$, where $\Delta^\I$ and $\Delta_{\D}$ are non-empty disjoint domains such that $d^\D\subseteq \Delta_{\D}$, for every $d \in N_{D}$, and $\cdot^\I$ is an interpretation function.
%, and of a \emph{concrete domain} $\Delta_{\D}$, disjoint from $\Delta^\I$, which is the domain %of interpretation of concrete datatypes. In particular, each datatype $d$ is interpreted as a non-empty subset c of $%\Delta_{\D}$. Moreover, it is required that the interpretations of the concrete datatypes form a partition of %$\Delta_{\D}$. 
%For space reasons, 
%The definition of 
The interpretation of concepts and roles, axioms and assertions is illustrated in Table \ref{semdlss}. 

{\small
\begin{longtable}{|>{\centering}m{2.5cm}|c|>{\centering\arraybackslash}m{6.7cm}|}
\hline
Name & Syntax & Semantics \\
\hline

concept & $A$ & $ A^\I \subseteq \Delta^\I$\\

ab. (resp., cn.) rl. & $R$ (resp., $P$ )& $R^\I \subseteq \Delta^\I \times \Delta^\I$ \hspace*{0.5cm} (resp., $P^\I \subseteq \Delta^\I \times \Delta_\D$)\\

%concrete role & $T$ & $T^\I \subseteq \Delta^\I \times \Delta_\D$\\

individual& $a$& $a^\I \in \Delta^\I$\\

nominal & $\{a\}$ & $\{a\}^\I = \{a^\I \}$\\

dtype  (resp., ng.) & $d$ (resp., $\neg d$)& $ d^\D \subseteq \Delta_\D$ (resp., $\Delta_\D \setminus d^\D $)\\

%data range $dr$ & $ dr $ & $  dr^{\D}\subseteq \Delta_{\D} $ \\

negative datatype term & $ \neg t_1 $ & $  (\neg t_1)^{\D} = \Delta_{\D} \setminus t_1^{\D}$ \\

datatype terms intersection & $ t_1 \sqcap t_2 $ & $  (t_1 \sqcap t_2)^{\D} = t_1^{\D} \cap t_2^{\D} $ \\

datatype terms union & $ t_1 \sqcup t_2 $ & $  (t_1 \sqcup t_2)^{\D} = t_1^{\D} \cup t_2^{\D} $ \\

constant in $N_{C}(d)$ & $ e_{d} $ & $ e_{d}^\D \in d^\D$ \\

%negated datatype & $\neg d$ & $\Delta_\D \setminus d^\D $\\

%const. in $N_{C}(d)$ & $e_{d}$ & $e_{d}^{\D} \in d^\D$ \\

\hline
data range  & $\{ e_{d_1}, \ldots , e_{d_n} \}$& $\{ e_{d_1}, \ldots , e_{d_n} \}^\D = \{e_{d_1}^\D \} \cup \ldots \cup \{e_{d_n}^\D \} $ \\

data range   &  $\psi_d$ & $\psi_d^\D$\\

data range    & $\neg dr$ &  $\Delta_\D \setminus dr^\D $\\

\hline

top (resp., bot.) & $\top$ (resp., $\bot$ )& $\Delta^\I$  (resp., $\emptyset$)\\

% & $\bot$ &  \\

negation & $\neg C$ & $(\neg C)^\I = \Delta^\I \setminus C$ \\

conj. (resp., disj.) & $C \sqcap D$ (resp., $C \sqcup D$)& $ (C \sqcap D)^\I = C^\I \cap D^\I$  (resp., $ (C \sqcup D)^\I = C^\I \cup D^\I$)\\

%disjunction & $C \sqcup D$ & $ (C \sqcup D)^\I = C^\I \cup D^\I$ \\

valued exist. quantification & $\exists R.{a}$ & $(\exists R.{a})^\I = \{ x \in \Delta^\I : \langle x,a^\I \rangle \in R^\I  \}$ \\

datatyped exist. quantif. & $\exists P.{e_{d}}$ & $(\exists P.e_{d})^\I = \{ x \in \Delta^\I : \langle x, e^\D_{d} \rangle \in P^\I  \}$ \\

self concept & $\exists R.\mathit{Self}$ & $(\exists R.\mathit{Self})^\I = \{ x \in \Delta^\I : \langle x,x \rangle \in R^\I  \}$ \\

nominals & $\{ a_1, \ldots , a_n \}$& $\{ a_1, \ldots , a_n \}^\I = \{a_1^\I \} \cup \ldots \cup \{a_n^\I \} $ \\

\hline

universal role & U & $(U)^\I = \Delta^\I \times \Delta^\I$\\

inverse role & $R^-$ & $(R^-)^\I = \{\langle y,x \rangle  \mid \langle x,y \rangle \in R^\I\}$\\

concept cart. prod. & $ C_1 \times C_2$   &  $ (C_1 \times C_2)^I = C_1^I \times C_2^I$ \\

abstract role complement & $ \neg R $ & $ (\neg R)^\I=(\Delta^\I \times \Delta^\I) \setminus R^\I $\\

abstract role union & $R_1 \sqcup R_2$ & $ (R_1 \sqcup R_2)^\I = R_1^\I \cup R_2^\I $\\

abstract role intersection & $R_1 \sqcap R_2$ & $ (R_1 \sqcap R_2)^\I = R_1^\I \cap R_2^\I $\\

abstract role domain restr. & $R_{C \mid }$ & $ (R_{C \mid })^\I = \{ \langle x,y \rangle \in R^\I : x \in C^\I  \} $\\

concrete role complement & $ \neg P $ & $ (\neg P)^\I=(\Delta^\I \times \Delta^\D) \setminus P^\I $\\

concrete role union & $P_1 \sqcup P_2$ & $ (P_1 \sqcup P_2)^\I = P_1^\I \cup P_2^\I $\\

concrete role intersection & $P_1 \sqcap P_2$ & $ (P_1 \sqcap P_2)^\I = P_1^\I \cap P_2^\I $\\

concrete role domain restr. & $P_{C \mid }$ & $ (P_{C \mid })^\I = \{ \langle x,y \rangle \in P^\I : x \in C^\I  \} $\\

concrete role range restr. & $P_{ \mid t}$ &  $ (P_{\mid t})^\I = \{ \langle x,y \rangle \in P^\I : y \in t^\D  \} $\\

concrete role restriction & $P_{ C_1 \mid t}$ &  $ (P_{C_1 \mid t})^\I = \{ \langle x,y \rangle \in P^\I : x \in C_1^\I \wedge y \in t^\D  \} $\\

\hline

concept subsum. & $C_1 \sqsubseteq C_2$ & $\I \models_\D C_1 \sqsubseteq C_2 \; \Longleftrightarrow \; C_1^\I \subseteq C_2^\I$ \\

ab. role subsum. & $ R_1 \sqsubseteq R_2$ & $\I \models_\D R_1 \sqsubseteq R_2 \; \Longleftrightarrow \; R_1^\I \subseteq R_2^\I$\\

role incl. axiom & $R_1 \ldots R_n \sqsubseteq R$ & $\I \models_\D R_1 \ldots R_n \sqsubseteq R  \; \Longleftrightarrow \; R_1^\I\circ \ldots \circ R_n^\I \subseteq R^\I$\\
cn. role subsum. & $ P_1 \sqsubseteq P_2$ & $\I \models_\D P_1 \sqsubseteq P_2 \; \Longleftrightarrow \; P_1^\I \subseteq P_2^\I$\\

\hline

symmetric role & $\sym(R)$ & $\I \models_\D \sym(R) \; \Longleftrightarrow \; (R^-)^\I \subseteq R^\I$\\

asymmetric role & $\asym(R)$ & $\I \models_\D \asym(R) \; \Longleftrightarrow \; R^\I \cap (R^-)^\I = \emptyset $\\

transitive role & $\tra(R)$ & $\I \models_\D \tra(R) \; \Longleftrightarrow \; R^\I \circ R^\I \subseteq R^\I$\\

disj. ab. role & $\mathsf{Dis}(R_1,R_2)$ & $\I \models_\D \mathsf{Dis}(R_1,R_2) \; \Longleftrightarrow \; R_1^\I \cap R_2^\I = \emptyset$\\

reflexive role & $\refl(R)$& $\I \models_\D \refl(R) \; \Longleftrightarrow \; \{ \langle x,x \rangle \mid x \in \Delta^\I\} \subseteq R^\I$\\

irreflexive role & $\irref(R)$& $\I \models_\D \irref(R) \; \Longleftrightarrow \; R^\I \cap \{ \langle x,x \rangle \mid x \in \Delta^\I\} = \emptyset  $\\

func. ab. role & $\fun(R)$ & $\I \models_\D \fun(R) \; \Longleftrightarrow \; (R^{-})^\I \circ R^\I \subseteq  \{ \langle x,x \rangle \mid x \in \Delta^\I\}$  \\

disj. cn. role & $\mathsf{Dis}(P_1,P_2)$ & $\I \models_\D \mathsf{Dis}(P_1,P_2) \; \Longleftrightarrow \; P_1^\I \cap P_2^\I = \emptyset$\\

func. cn. role & $\fun(P)$ & $\I \models_\D \fun(p) \; \Longleftrightarrow \; \langle x,y \rangle \in P^\I \mbox{ and } \langle x,z \rangle \in P^\I \mbox{ imply } y = z$  \\

\hline

datatype terms equivalence & $ t_1 \equiv t_2 $ & $ \I \models_{\D} t_1 \equiv t_2 \Longleftrightarrow t_1^{\D} = t_2^{\D}$\\

datatype terms diseq. & $ t_1 \not\equiv t_2 $ & $ \I \models_{\D} t_1 \not\equiv t_2 \Longleftrightarrow t_1^{\D} \neq t_2^{\D}$\\

datatype terms subsum. & $ t_1 \sqsubseteq t_2 $ &  $ \I \models_{\D} (t_1 \sqsubseteq t_2) \Longleftrightarrow t_1^{\D} \subseteq t_2^{\D} $ \\

\hline

concept assertion & $a : C_1$ & $\I \models_\D a : C_1 \; \Longleftrightarrow \; (a^\I \in C_1^\I) $ \\

agreement & $a=b$ & $\I \models_\D a=b \; \Longleftrightarrow \; a^\I=b^\I$\\

disagreement & $a \neq b$ & $\I \models_\D a \neq b  \; \Longleftrightarrow \; \neg (a^\I = b^\I)$\\

%%datatype agreement & $ d_1=d_2$ & $ d_{1}^\D = d_{2}^\D$\\
%%
%%datatype disagreement & $ d_{1} \neq d_{2}$ & $\neg (d_{1}^\D = d_{2}^\D)$\\

ab. role asser. & $ (a,b) : R $ & $\I \models_\D (a,b) : R \; \Longleftrightarrow \;  \langle a^\I , b^\I \rangle \in R^\I$ \\

cn. role asser. & $ (a,e_d) : P $ & $\I \models_\D (a,e_d) : P \; \Longleftrightarrow \;   \langle a^\I , e_d^\D \rangle \in P^\I$ \\

\hline \caption{Semantics of $\shdlssx$.}\\
\caption*{\emph{Legenda.} ab: abstract, cn.: concrete, rl.: role, ind.: individual, d. cs.: datatype constant, dtype: datatype, ng.: negated, bot.: bottom, incl.: inclusion, asser.: assertion.}  \label{semdlss}
\end{longtable}}

%\vspace{-0.3cm}

Let $\mathcal{KB}=(\mathcal{R}, \mathcal{T}, \mathcal{A})$ be a $\shdlssx$-knowledge base. An interpretation $\I= (\Delta ^ \I, \Delta_{\D}, \cdot ^ \I)$ is a $\D$-model of $\mathcal{R}$ (and  write $\I \models_{\D} \mathcal{R}$) if $\I$ satisfies each axiom in $\mathcal{R}$ according to the semantic rules in \cite[Table~1]{ExtendedVersionICTCS2016}. Similar definitions hold for $\mathcal{T}$ and $\mathcal{A}$ too. Then $\I$ satisfies $\mathcal{KB}$ (and write $\I \models_{\D} \mathcal{KB}$) if it is a $\D$-model of $\mathcal{R}$, $\mathcal{T}$, and $\mathcal{A}$. A knowledge base is \emph{consistent} if it is satisfied by some interpretation.

\section{Conjunctive Query Answering for $\shdlssx$}
%We now introduce formally the problem of CQA.
Let $ \mathcal{V} = \{v_{1}, v_{2}, \ldots\}$ be a denumerable and infinite set of variables disjoint from $\Ind$ and from $\bigcup\{N_C(d): d \in N_{\D}\}$. A  $\shdlssx$-\emph{atomic formula} is an expression of of the following types\\[.3cm]
\centerline{$R(w_1,w_2)$, \quad $P(w_1, u_1)$, \quad $C(w_1)$, \quad $w_1=w_2$, \quad $u_1 = u_2$,}\\[.3cm]
where $w_1,w_2 \in \mathcal{V}\cup \Ind$, $u_1, u_2 \in  \mathcal{V}\cup \bigcup \{N_C(d): d \in N_{\D}\}$, $R$ is a $\shdlssx$-abstract role term, $P$ is a $\shdlssx$-concrete role term, and $C$ is a $\shdlssx$-concept term. A $\shdlssx$-atomic formula containing no variables is said to be \emph{closed}. A $\shdlssx$-\emph{literal} is a $\shdlssx$-atomic formula or its negation. 
%A clause is a finite disjunction of literals. 
 A $\shdlssx$-\emph{conjunctive query} is a conjunction of $\shdlssx$-literals. 
Let  $v_1,\ldots,v_n \in \mathcal{V}$ and $o_1, \ldots, o_n \in \Ind \cup \bigcup \{N_C(d): d \in N_{\D}\}$. A substitution $\sigma  \defAs \{v_1/o_1, \ldots, v_n/o_n \}$ is a map such that, for every  $\shdlssx$-literal $L$, $L\sigma$ is obtained from $L$ by replacing the occurrences of $v_1, \ldots, v_n$ in $L$ with $o_1, \ldots, o_n$, respectively. Substitutions can be extended to $\shdlssx$-conjunctive queries in the usual way. 
%$T_1 \wedge \ldots \wedge T_n$ is defined as follows: $(T_1 \wedge \ldots \wedge T_m)\sigma =_{Def} T_1 \sigma \wedge \ldots \wedge T_m \sigma$.
Let $Q \defAs  (L_1 \wedge \ldots \wedge L_m)$ be a $\shdlssx$-conjunctive query, and $\KB$ a $\shdlssx$-knowledge base. A substitution $\sigma$ involving \emph{exactly} the variables occurring in $Q$ is a \emph{solution for $Q$ w.r.t. $\KB$} if there exists a $\shdlssx$-interpretation $\I$ such that $\I \models_{\D} \KB$ and $\I \models_{\D} Q \sigma$. The collection $\Sigma$ of the  solutions for $Q$ w.r.t. $\KB$ is the \emph{answer set of $Q$ w.r.t. $\KB$}. Then the \emph{conjuntive query answering} (CQA) problem for $Q$ w.r.t. $\KB$ consists in finding the answer set $\Sigma$ of $Q$ w.r.t. $\KB$. 

We shall solve the CQA problem just stated by reducing it to the analogous problem formulated in the context of the fragment $\flqsr$ (and in turn to the decision procedure for $\flqsr$ presented in \cite{CanNic2013}). The CQA problem for $\flqsr$-formulae can be stated as follows.
%Before addressing the CQA problem for $\shdlssx$, it is useful to consider first the same problem in the context of $\flqsr$-formulae.  
%the \textit{conjunctive query answering} problem in the context of the $\flqsr$-fragment.
Let $\phi$ be a $\flqsr$-formula and let $\psi$ be a conjunction of $\flqsr$-quantifier-free atomic formulae of level $0$ of the types\\[0.3cm] 
\centerline{$x=y$,\qquad $x \in X^1$, \qquad $ \langle x,y \rangle \in X^3$}\\[.3cm] 
or their negations, such that $\varz(\psi) \cap \varz(\phi)=\emptyset$ and 
$\varu(\psi) \cup \vart(\psi) \subseteq \varu(\phi) \cup \vart(\phi)$. 
The \emph{CQA problem for $\psi$ w.r.t.\ $\phi$} consists in computing the \emph{answer set of $\psi$ w.r.t.\ $\phi$}, namely the collection $\Sigma'$ of all the  substitutions $\sigma' \defAs \{x_1 / y_1, \ldots, x_n / y_n \}$ (where $x_1, \ldots, x_n$ are the distinct variables of level 0 in $\psi$ and  $\{y_1, \ldots, y_n\} \subseteq \varz(\phi)$) such that $\M \models \phi \wedge \psi\sigma'$, for some $\flqsr$-interpretation $\M$. 
%Plainly, $\Sigma' = \bigcup_{\M \models \phi}  \Sigma'_{\M}$, where $\Sigma'_{\M}$ is the collection of all the substitutions $\sigma$ satisfying the above requirement such that $\M \models \phi \wedge \psi\sigma$.
%
%$\flqsr$ consists, for all $\flqsr$-formulae $\phi$ and $\psi$ satisfying the  above requirements, in determining, for every $\flqsr$-interpretation $\M$ such that $\M \models \phi$, the collection $\Sigma'_{\M}$ of all the substitutions $\sigma \defAs \{x_1 / y_1, \ldots, x_n / y_n \}$, where $x_1, \ldots, x_n$ are the distinct variables of level 0 in $\psi$ and  $\{y_1, \ldots, y_n\} \subseteq \varz(\phi)$, such that  $\M\models  \psi\sigma$, and then in calculating the set $\Sigma'$ such that 
%%$\Sigma' \defAs \overset{}{\underset{\M \models \phi}{\bigcup}}  \Sigma'_{\M}$.
%$\Sigma' \defAs \bigcup_{\M \models \phi}  \Sigma'_{\M}$.
In view of the decidability of the satisfiability problem for $\flqsr$-formulae, the CQA problem for $\flqsr$-formulae can be solved effectively. Indeed, given two $\flqsr$-formulae $\phi$ and $\psi$ satisfying the above requirements, to compute the answer set of $\psi$ w.r.t.\ $\phi$, for each candidate substitution $\sigma' \defAs \{x_1 / y_1, \ldots, x_n / y_n \}$ (with $\{x_1, \ldots, x_n \}=  \varz (\psi) $ and  $\{y_1, \ldots, y_n\} \subseteq \varz(\phi)$) one has just to test for satisfiability the $\flqsr$-formula $\phi \wedge \psi\sigma'$. Since the number of possible candidate substitutions is $|\varz (\phi)|^{|\varz (\psi)|}$ and the satisfiability test for $\flqsr$-formulae can be carried out in an effective manner, the answer set of $\psi$ w.r.t.\ $\phi$ can be computed effectively. Summarizing,
\begin{lemma}\label{CQA4LQSR}
The CQA problem for $\flqsr$-formulae can be solved in an effective manner. \qed
\end{lemma}

The following theorem states that also the CQA problem for $\shdlssx$ can be solved effectively. 
\begin{theorem}\label{CQADL}
Given a $\shdlssx$-knowledge base $\KB$ and a $\shdlssx$-conjunc\-tive query $Q$, the CQA problem for $Q$ w.r.t. $\KB$ can be solved in an effective manner.
\end{theorem}
%
%\begin{proof}[sketch]
We first outline the main ideas and then we provide a formal proof of the theorem. 
%of the proof, th The interested reader, however, can find complete details in the extended version of this paper (see \cite{ExtendedVersionICTCS2016}).

As remarked above, the CQA problem for $\shdlssx$ can be solved via an effective reduction to the CQA problem for $\flqsr$-formulae, and then exploiting Lemma \ref{CQA4LQSR}. The reduction is accomplished through a function $\theta$ that maps effectively variables in $\mathcal{V}$, individuals in $\Ind$, datatype constants in $\bigcup\{N_C(d): d \in N_{\D}\}$ into variables of sort 0 (of the $\flqsr$-language), etc., $\shdlssx$\emph{-TBoxes}, \emph{-RBoxes}, and \emph{-ABoxes}, and $\shdlssx$-conjunctive queries into $\flqsr$-formulae in conjunctive normal form (CNF), which can be used to map effectively CQA problems from the $\shdlssx$-context into the $\flqsr$-context. More specifically, given a $\shdlssx$-knowledge base $\KB$ and a $\shdlssx$-conjunctive query $Q$, using the function $\theta$ we can effectively construct the following $\flqsr$-formulae in CNF:\\[0.2cm]
\centerline{
$\phi_{\KB} \defAs \bigwedge_{H \in \KB} \theta(H) \wedge \bigwedge_{i=1}^{12} \xi_i, \qquad \psi_Q \defAs \theta(Q)\,.$\footnotemark
}\\[.2cm]
Then, if we denote by $\Sigma$ the answer set of $Q$ w.r.t.\ $\KB$ and by $\Sigma'$ the answer set of $\psi_Q$ w.r.t.\ $\phi_{\KB}$, we have that $\Sigma$ consists of all substitutions $\sigma$ (involving exactly the variables occurring in $Q$) such that $\theta(\sigma) \in \Sigma'$.
Since, by Lemma~\ref{CQA4LQSR}, $\Sigma'$ can be computed effectively, then $\Sigma$ can be computed effectively too.

\footnotetext{The definition of the function $\theta$ is inspired to that of the function $\tau$ introduced in the proof of Theorem 1 in \cite{CanLonNicSanRR2015}. Specifically, $\theta$ differs from $\tau$ as (i) it allows quantification only on variables of level $0$, (ii) it treats Boolean operations on concrete roles and the product of concepts, and (iii) it constructs $\flqsr$-formulae in CNF. In addition, the constraints $\xi_1$--$\xi_{12}$ are similar to the constraints $\psi_1$--$\psi_{12}$ introduced in the proof of Theorem 1 in \cite{CanLonNicSanRR2015}; they are introduced to guarantee that each model of $\phi_{\KB}$ can be transformed into a $\shdlssx$-interpretation. }
We are now ready to provide the proof of Theorem 1.
\begin{proof}
As preliminary step, observe that the statements of $\KB$ that need to be considered are the following:
\begin{itemize}[itemsep=0.1cm]
\item[-] $C_1 \equiv \top$, $C_1 \equiv \neg C_2$, $C_1 \equiv C_2 \sqcup C_3$, $C_1 \equiv \{a\}$, $C_1 \sqsubseteq \forall R_1.C_2$, $\exists R_1.C_1 \sqsubseteq C_2$, $\geq_n\!\! R_1. C_1 \sqsubseteq C_2$, $C_1 \sqsubseteq {\leq_n\!\! R_1. C_2}$, $C_1 \sqsubseteq \forall P_1.t_1$, $\exists P_1.t_1 \sqsubseteq C_1$, $\geq_n\!\! P_1. t_1 \sqsubseteq C_1$, $C_1 \sqsubseteq {\leq_n\!\! P_1. t_1}$,
\item[-] $R_1 \equiv U$, $R_1 \equiv \neg R_2$, $R_1 \equiv R_2 \sqcup R_3$, $R_1 \equiv R_2^{-}$, $R_1 \equiv id(C_1)$, $R_1 \equiv R_{2_{C_1 |}}$, $R_1 \ldots R_n \sqsubseteq R_{n+1}$, $\refl(R_1)$, $\irref(R_1)$, $\mathsf{Dis}(R_1,R_2)$, $\fun(R_1)$, $R_1 \equiv C_1 \times C_2$,
\item[-] $P_1 \equiv P_2$, $P_1 \equiv \neg P_2$, $P_1 \equiv P_2 \sqcup P_3$, $P_1 \sqsubseteq P_2$, $\fun(P_1)$, $P_1 \equiv P_{2_{C_1 |}}$, $P_1 \equiv P_{2_{C_1 | t_1}}$, $P_1 \equiv P_{2_{| t_1}}$,
%\item[-] $t_1 \equiv t_2$, $t_1 \equiv \neg t_2$, $t_1 \equiv t_2 \sqcup t_3$, $t_1 \equiv \{e_d\}$,
\item[-] $a : C_1$, $(a,b) : R_1$, $(a,b) : \neg R_1$, $a=b$, $a \neq b$,
\item
$e_{d} : t_1$, $(a, e_{d}) : P_1$, $(a, e_{d}) : \neg P_1$.
\end{itemize}

 We solve the problem of CQA for $\shdlssx$ via a reduction to the problem of CQA for 
$\flqsr$, exploiting the decidability result  proved in Lemma \ref{CQA4LQSR}.

%Specifically we define a mapping $\theta$ such that for every (minimal) $\sigma$, $\KB \models_D Q\sigma$ iff $\varphi_{\KB} \models \psi_Q \sigma'$, where $\varphi_{\KB} = \underset{H \in \KB}{\theta(H)} \wedge \overset{12}{\underset{i=1}{\bigwedge}} \xi_i$, $\psi_Q = \theta(Q)$, and $\sigma'=\theta(\sigma)$. The definition of the mapping $\theta$, reported in Appendix \ref{apptheta}, is inspired to the definition of the mapping $\tau$, given in \cite{CanLonNicSan15}. $\theta$ differs from $\tau$ because it yields $\flqsr$-formulae in conjunctive normal form that admit quantifiers only on variables of level $0$. Moreover $\theta$ threats boolean operations on concrete roles and the product of concepts.

%The constraints $\xi_1-\xi_12$, defined in Appendix \ref{apptheta} are introduced to guarantee that each model of $\varphi_{\KB}$ can be easily transformed in a $\shdlssx$-interpretation.
%

We define  
a function $\theta$ that maps the $\shdlssx$-knowledge base $\KB$ and the $\shdlssx$-conjunctive query $Q$ in the $\flqsr$-formulae in Conjunctive Normal Form (CNF) $\phi_{\KB}$ and $\psi_{Q}$, respectively, and the answer set $\Sigma$ for $Q$ w.r.t. $\KB$ in a set $\Sigma'$ of (0 level) substitutions in the $\flqsr$ formalism.

We will show that, $\Sigma$ is the answer set for $Q$ w.r.t. $\KB$ iff $\Sigma$ is equal to $\Sigma' = \overset{}{\underset{\M \models \phi_{\KB}}{\bigcup}}  \Sigma_{\M}'$, where $\Sigma_{\M}'$ is the collection of substitutions $\sigma'$ such that $\M \models \psi_{Q}\sigma'$. 

The definition of the mapping $\theta$ is inspired to the definition of the mapping $\tau$ introduced in the proof of Theorem 1 in \cite{CanLonNicSanRR2015}. Specifically, $\theta$ differs from $\tau$ because it allows quantification only on variables of level $0$, it treats Boolean operations on concrete roles and the product of concepts, and it construct $\flqsr$-formulae in CNF.
 To prepare for the definition of $\theta$, we map injectively individuals $a$, constants $e_d \in N_{C}(d)$, and variable $y,z,\ldots \in \mathcal{V}$, into level $0$ variables $x_a$, $x_{e_d}$, $x_{y}$, $x_{z}$, the constant concepts $\top$ and $\bot$, datatype terms $t$, and concept terms $C$ into level $1$ variables $X_{\top}^1$, $X_{\bot}^1$, $X_{t}^1$, $X_{C}^1$, respectively, and the universal relation on individuals $U$, abstract role terms $R$, and concrete role terms $P$ into level $3$ variables $X_{U}^3$, $X_{R}^3$, and $X_{P}^3$, respectively.\footnote{The use of level $3$ variables to model abstract and concrete role terms is motivated by the fact that their elements, that is ordered pairs $\langle x, y \rangle$, are encoded in Kuratowski's style as $\{\{x\}, \{x,y\}\}$, namely as collections of sets of objects.}

Then the mapping $\theta$ is defined as follows:
\smallskip

\noindent $\theta(C_1 \equiv \top) \defAs (\forall z)( ( \neg(z \in X_{C_1}^1) \vee z \in X_{\top}^1) \wedge ( \neg(z \in X_{\top}^1) \vee z \in X_{C_1}^1))$,

\noindent $\theta(C_1 \equiv \neg C_2) \defAs (\forall z)(( \neg(z \in X_{C_1}^1) \vee \neg(z \in X_{C_2}^1)) \wedge (z \in X_{C_2}^1 \vee z \in X_{C_1}^1))$,

\noindent $\theta(C_1 \equiv C_2 \sqcup C_3 ) \defAs (\forall z)( ( \neg(z \in X_{C_1}^1) \vee (z \in X_{C_2}^1 \vee z \in X_{C_3}^1)) \wedge ( (\neg (z \in X_{C_2}^1) \vee z \in X_{C_1}^1) \wedge (\neg (z \in X_{C_3}^1) \vee z \in X_{C_1}^1 ))$,

\noindent $\theta(C_1 \equiv \{a\}) \defAs (\forall z)( \neg(z \in X_{C_1}^1) \vee z = x_a) \wedge( \neg(z = x_a) \vee z \in X_{C_1}^1 )$,

\noindent $\theta(C_1 \sqsubseteq \forall R_1.C_2) \defAs (\forall z_1)(\forall z_2)( \neg(z_1 \in X_{C_1}^1) \vee ( \neg(\langle z_1,z_2 \rangle \in X_{R_1}^3) \vee z_2 \in X_{C_2}^1))$,

\noindent $\theta(\exists R_1.C_1 \sqsubseteq C_2) \defAs (\forall z_1)(\forall z_2)(( \neg(\langle z_1,z_2 \rangle \in X_{R_1}^3) \vee \neg( z_2 \in X_{C_1}^1)) \vee z_1 \in X_{C_2}^1)$,

\noindent $\theta(C_1 \equiv \exists R_1.\{a\}) \defAs(\forall z)( ( \neg(z \in X_{C_1}^1) \vee \langle z,x_{a}\rangle \in X_{R_1}^3) \wedge ( \neg(\langle z,x_{a}\rangle \in X_{R_1}^3) \vee z \in X_{C_1}^1  ) )$,

\noindent $\theta(C_1 \sqsubseteq \leq_n\!\! R_1.C_2) \defAs (\forall z)(\forall z_1)\ldots (\forall z_{n+1})( \neg(z \in X_{C_1}^1) \vee  ( \overset{n+1}{\underset{i=1}\bigwedge}( \neg(z_i \in X_{C_2}) \vee \neg(\langle z,z_i\rangle \in X_{R_1}^3) \vee \underset {i<j} {\bigvee} z_i = z_j))$,

\noindent $\theta(\geq_n\!\! R_1.C_1 \sqsubseteq C_2) \defAs (\forall z)(\forall z_1)\ldots (\forall z_{n})( \overset {n}{ \underset{i=1}\bigwedge }(( \neg(z_i \in X_{C_1}^1) \vee \neg( \langle z,z_i\rangle \in X_{R_1}^3)) \vee  \underset {i<j} \bigvee z_i = z_j) \vee z \in X_{C_2}^1)$,

\noindent $\theta(C_1 \sqsubseteq \forall P_1.t_1) \defAs (\forall z_1)(\forall z_2)( \neg(z_1 \in X_{C_1}^1) \vee ( \neg (\langle z_1,z_2 \rangle \in X_{P_1}^3) \vee z_2 \in X_{t_1}^1))$,

\noindent $\theta(\exists P_1.t_1 \sqsubseteq C_1) \defAs (\forall z_1)(\forall z_2)(( \neg(\langle z_1,z_2 \rangle \in X_{P_1}^3) \vee \neg(z_2 \in X_{t_1}^1)) \vee z_1 \in X_{C_1}^1)$,

\noindent $\theta(C_1 \equiv \exists P_1.\{e_{d}\}) \defAs (\forall z)( ( \neg(z \in X_{C_1}^1) \vee \langle z,x_{e_{d}}\rangle \in X_{P_1}^3)  \wedge ( \neg(\langle z,x_{e_{d}}\rangle \in X_{P_1}^3) \vee z \in X_{C_1}^1) )$,

\noindent $\theta(C_1 \sqsubseteq \leq_n\!\! P_1.t_1) \defAs (\forall z)(\forall z_1)\ldots (\forall z_{n+1})( \neg (z \in X_{C_1}^1) \vee ( \overset {n+1} { \underset{i=1}\bigwedge }( \neg(z_i \in X_{t_1}) \vee \neg(\langle z,z_i\rangle \in X_{P_1}^3) \vee \underset{i<j} {\bigvee} z_i = z_j))$,

\noindent $\theta(\geq_n\!\! P_1.t_1 \sqsubseteq C_1) \defAs (\forall z)(\forall z_1)\ldots (\forall z_{n})( \overset {n} { \underset {i=1}\bigwedge}(( \neg(z_i \in X_{t_1}^1) \vee \neg(\langle z,z_i\rangle \in X_{P_1}^3)) \vee \underset {i<j} {\bigvee} z_i = z_j) \vee z \in X_{C_1}^1)$,

\noindent $\theta(R_1 \equiv U) \defAs (\forall z_1)(\forall z_2)( ( \neg(\langle z_1,z_2\rangle \in X_{R_1}^3) \vee \langle z_1,z_2\rangle \in X_{U}^3) \wedge ( \neg(\langle z_1,z_2\rangle \in X_{U}^3) \vee \langle z_1,z_2\rangle \in X_{R_1}^3) )$,

\noindent $\theta(R_1 \equiv \neg R_2) \defAs (\forall z_1)(\forall z_2)( ( \neg(\langle z_1,z_2\rangle \in X_{R_1}^3) \vee \neg (\langle z_1,z_2\rangle \in X_{R_2}^3 )) \wedge ( \langle z_1,z_2\rangle \in X_{R_2}^3 \vee \neg (\langle z_1,z_2\rangle \in X_{R_1}^3 )) )$,

\noindent $\theta( R \equiv C_1 \times C_2 ) \defAs (\forall z_1)(\forall z_2) (  \neg (\langle z_1, z_2 \rangle \in X^3_R) \vee  z_1 \in X^1_{C_1}) \wedge ( \neg (\langle z_1, z_2 \rangle \in X^3_R) \vee  z_2  \in X^1_{C_2}  ) \wedge  (( \neg(z_1 \in X^1_{C_1}) \vee  \neg(z_2  \in X^1_{C_2})  ) \vee \langle z_1, z_2 \rangle \in X^3_R  ) )$

\noindent $\theta(R_1 \equiv R_2 \sqcup R_3) \defAs (\forall z_1)(\forall z_2)( ( \neg(\langle z_1,z_2 \rangle \in X_{R_1}^3) \vee (\langle z_1,z_2 \rangle \in X_{R_2}^3 \vee \langle z_1,z_2 \rangle \in X_{R_3}^3)) \wedge ( ( \neg(\langle z_1,z_2 \rangle \in X_{R_2}^3) \vee \langle z_1,z_2 \rangle \in X_{R_1}^3) \wedge ((\neg(\langle z_1,z_2 \rangle \in X_{R_3}^3) \vee \langle z_1,z_2 \rangle \in X_{R_1}^3) ) ))$,

\noindent $\theta(R_1 \equiv R_2^{-}) \defAs (\forall z_1)(\forall z_2)( ( \neg(\langle z_1,z_2\rangle \in X_{R_1}^3) \vee \langle z_2,z_1\rangle \in X_{R_2}^3 ) \wedge ( \neg(\langle z_2,z_1\rangle \in X_{R_2}^3) \vee \langle z_1,z_2\rangle \in X_{R_1}^3  )   )  $,

\noindent $\theta(R_1 \equiv id(C_1)) \defAs (\forall z_1)(\forall z_2)( ( ( \neg(\langle z_1,z_2\rangle \in X_{R_1}^3) \vee z_1 \in X_{C_1}^1 ) \wedge ( \neg(\langle z_1,z_2\rangle \in X_{R_1}^3) \vee z_2 \in X_{C_1}^1 ) \wedge (\neg(\langle z_1,z_2\rangle \in X_{R_1}^3) \vee z_1 =z_2)  )\wedge ( ( \neg(z_1 \in X_{C_1}^1) \vee \neg(z_2 \in X_{C_1}^1) \vee z_1 \neq z_2) \vee \langle z_1,z_2\rangle \in X_{R_1}^3)  )$,

\noindent $\theta(R_1 \equiv R_{2_{C_1 |}}) \defAs (\forall z_1)(\forall z_2)( ( (\neg(\langle z_1,z_2\rangle \in X_{R_1}^3) \vee \langle z_1,z_2\rangle \in X_{R_2}^3) \wedge  ( \neg(\langle z_1,z_2\rangle \in X_{R_1}^3) \vee z_1 \in X_{C_1}^1))  \wedge (( \neg(\langle z_1,z_2\rangle \in X_{R_2}^3) \vee \neg(z_1 \in X_{C_1}^1))  \vee \langle z_1,z_2\rangle \in X_{R_1}^3 ) )$,

\noindent $\theta(R_1 \ldots R_n \sqsubseteq R_{n+1}) \defAs (\forall z)(\forall z_1)\ldots (\forall z_{n}) (( \neg(\langle z, z_1\rangle \in X_{R_1}^3) \vee \ldots \vee \neg(\langle z_{n-1},z_n\rangle \in X_{R_{n}}^3)) \vee  \langle z, z_n\rangle\in X_{R_{n+1}}^3)$,

\noindent $\theta(\refl(R_1)) \defAs (\forall z)(\langle z, z\rangle \in X_{R_1}^3)$,

\noindent $\theta(\irref(R_1)) \defAs (\forall z)(\neg (\langle z,z\rangle \in X_{R_1}^3))$,

\noindent $\theta(\fun(R_1)) \defAs (\forall z_1)(\forall z_2)(\forall z_3)(( \neg(\langle z_1,z_2\rangle \in X_{R_1}^3) \vee \neg(\langle z_1,z_3\rangle \in X_{R_1}^3)) \vee z_2 =z_3)$,

\noindent $\theta(P_1 \equiv P_2) \defAs (\forall z_1)(\forall z_2)( ( \neg (\langle z_1,z_2\rangle \in X_{P_1}^3) \vee \langle z_1,z_2\rangle \in X_{P_2}^3) \wedge ( \neg(\langle z_1,z_2\rangle \in X_{P_2}^3) \vee  \langle z_1,z_2\rangle \in X_{P_1}^3)  )$,

\noindent $\theta(P_1 \equiv \neg P_2) \defAs (\forall z_1)(\forall z_2)( ( \neg(\langle z_1,z_2\rangle \in X_{P_1}^3) \vee \neg(\langle z_1,z_2\rangle \in X_{P_2}^3)) \wedge (\langle z_1,z_2\rangle \in X_{P_2}^3 \vee \langle z_1,z_2\rangle \in X_{P_1}^3 ) )$,

\noindent $\theta(P_1 \sqsubseteq P_2) \defAs (\forall z_1)(\forall z_2)( \neg(\langle z_1,z_2\rangle \in X_{P_1}^3) \vee \langle z_1,z_2\rangle \in X_{P_2}^3)$,

\noindent $\theta(\fun(P_1)) \defAs (\forall z_1)(\forall z_2)(\forall z_3)( ( \neg(\langle z_1,z_2\rangle \in X_{P_1}^3) \vee \neg(\langle z_1,z_3\rangle \in X_{P_1}^3) \vee z_2 =z_3)$,

\noindent $\theta(P_1 \equiv P_{2_{C_1 |}}) \defAs (\forall z_1)(\forall z_2) (( \neg(\langle z_1,z_2\rangle \in X_{P_1}^3) \vee \langle z_1,z_2\rangle \in X_{P_2}^3 ) \wedge ( \neg(\langle z_1,z_2\rangle \in X_{P_1}^3) \vee z_1 \in X_{C_1}^1) \wedge ( (\neg \langle z_1,z_2\rangle \in X_{P_2}^3) \vee \neg(z_1 \in X_{C_1}^1) \vee  \langle z_1,z_2\rangle \in X_{P_1}^3) )$,

\noindent $\theta(P_1 \equiv P_{2_{|t_1}}) \defAs (\forall z_1)(\forall z_2) ( ( \neg(\langle z_1,z_2\rangle \in X_{P_1}^3) \vee \langle z_1,z_2\rangle \in X_{P_2}^3 ) \wedge ( \neg(\langle z_1,z_2\rangle \in X_{P_1}^3) \vee z_2 \in X_{t_1}^1) \wedge ( ( \neg(\langle z_1,z_2\rangle \in X_{P_2}^3) \vee \neg(z_2 \in X_{t_1}^1)) \vee \langle z_1,z_2\rangle \in X_{P_1}^3 ) )$,

\noindent $\theta(P_1 \equiv P_{2_{C_1|t_1}}) \defAs (\forall z_1)(\forall z_2) ( ( \neg(\langle z_1,z_2\rangle \in X_{P_1}^3) \vee \langle z_1,z_2\rangle \in X_{P_2}^3 ) \wedge ( \neg(\langle z_1,z_2\rangle \in X_{P_1}^3) \vee z_1 \in X_{C_1}^1) \wedge ( \neg(\langle z_1,z_2\rangle \in X_{P_1}^3) \vee z_2 \in X_{t_1}^1)
 \wedge (    \neg(\langle z_1,z_2\rangle \in X_{P_2}^3) \vee \neg(z_1 \in X_{C_1}^1) \vee \neg(z_2 \in X_{t_1}^1)  \vee \langle z_1,z_2\rangle \in X_{P_1}^3) )$,

\noindent $\theta(t_1 \equiv t_2)\defAs (\forall z)(( \neg(z \in X_{t_1}^1) \vee z \in X_{t_2}^1) \wedge ( \neg(z \in X_{t_2}^1) \vee z \in X_{t_1}^1 ))$,
\noindent $\theta(t_1 \equiv \neg t_2)\defAs (\forall z)(( \neg(z \in X_{t_1}^1) \vee \neg (z \in X_{t_2}^1)) \wedge ( z \in X_{t_2}^1  \vee z \in X_{t_1}^1) )$,

\noindent $\theta(t_1 \equiv t_2 \sqcup t_3)\defAs (\forall z)( ( \neg(z \in X_{t_1}^1) \vee (z \in X_{t_2}^1\vee z \in X_{t_3}^1)) \wedge ( ( \neg (z \in X_{t_2}^1) \vee  z \in X_{t_1}^1) \wedge (\neg (z \in X_{t_3}^1) \vee  z \in X_{t_1}^1) ) )$,

\noindent $\theta(t_1 \equiv t_2 \sqcap t_3)\defAs (\forall z)(( \neg(z \in X_{t_1}^1) \vee (z \in X_{t_2}^1 \wedge z \in X_{t_3}^1))  \wedge ((( \neg(z \in X_{t_2}^1) \vee \neg(z \in X_{t_3}^1))  \vee z \in X_{t_1}^1)  )$,

\noindent $\theta(t_1 \equiv \{e_{d}\})\defAs (\forall z)((\neg (z \in X_{t_1}^1) \vee z = x_{e_{d}}) \wedge ( \neg(z = x_{e_{d}})  \vee z \in X_{t_1}^1) )$,

\noindent $\theta(a : C_1) \defAs x_a \in X_{C_1}^1$,

\noindent $\theta((a,b) : R_1) \defAs \langle x_a, x_b\rangle \in X_{R_1}^3$,

\noindent $\theta((a,b) : \neg R_1) \defAs \neg(\langle x_a, x_b\rangle \in X_{R_1}^3)$,

\noindent $\theta(a=b) \defAs x_a = x_b$, $\theta(a\neq b) \defAs \neg (x_a = x_b)$,

\noindent $\theta(e_d : t_1) \defAs x_{e_d} \in X_{t_1}^1$,

\noindent $\theta((a,e_d) : P_1) \defAs \langle x_a, x_{e_d}\rangle \in X_{P_1}^3$, $\theta((a,e_d) : \neg P_1) \defAs \neg(\langle x_a, x_{e_d}\rangle \in X_{P_1}^3)$,

\noindent $\theta(\alpha \wedge \beta) \defAs \theta(\alpha) \wedge \theta(\beta)$.

\medskip \noindent The mapping $\theta$ for $\shdlssx$-conjuctive queries is defined as follows.\\

\noindent$\theta (R_1 (w_1,w_2)) \defAs \langle x_{w_1}, x_{w_2} \rangle \in X^3_{R_1}$ ,\\
$\theta (P_1 (w_1,u_1)) \defAs \langle x_{w_1}, x_{u_1} \rangle \in X^3_{P_1}$,\\
$\theta ( C_1(w_1) \defAs x_{w_1} \in X^1_{C_1}$, \\
$\theta (w_1 =w_2) \defAs x_{w_1} = x_{w_2}$, \\
$\theta (u_1 =u_2) \defAs x_{u_1} = x_{u_2}$.\\
%$\theta ( \alpha_x \wedge \beta_y) \defAs \theta(\alpha_x) \wedge \theta(\beta_y) $, where $x$ occurs in $\alpha$ and $y$ in $\beta$,
%\\
%$\theta ( \alpha_x \vee \beta_y) \defAs \theta(\alpha_x) \vee \theta(\beta_y) $, where $x$ occurs in $\alpha$ and $y$ in $\beta$,
\\
%$\theta ( \neg \alpha_x) \defAs \neg \theta(\alpha_x)$, where $x$ occurs in $\alpha$.\\

To complete, we extend the mapping $\theta$ on substitutions $\sigma \defAs \{x_1/o_1,\ldots,x_n/o_n\}$, where $x_1,\ldots x_n \in \mathcal{V}$ and $o_1,\ldots,o_n \in \Ind \cup \bigcup\{N_C(d): d \in N_{\D}\}$. 

We put $\theta(\sigma)$= $\theta(\{x_1/o_1,\ldots,x_n/o_n\})$ = $\{x_{x_1}/x_{o_1}, \ldots,  x_{x_n}/x_{o_n}\} = \sigma'$, where $x_{x_1}, \ldots, {x_n},$ $x_{o_1}, \ldots, x_{o_n}$ are variables of level $0$ in $\flqsr$.

Let $\mathcal{KB}$ be our $\shdlssx$-knowledge base, and let $\ck$, $\ark$, $\crk$, and $\ik$ be, respectively, the sets of concept, of abstract role, of concrete role, and of individual names in $\mathcal{KB}$. Moreover, let $N_{D}^\mathcal{KB} \subseteq N_{D}$ be the set of datatypes in $\mathcal{KB}$, $N_{F}^\mathcal{KB}$ a restriction of $N_{F}$ assigning to every $d \in N_{\D}^\mathcal{KB}$ the set $N_{F}^\mathcal{KB}(d)$ of facets in $N_{F}(d)$ and in $\mathcal{KB}$. Analogously, let $N_{C}^{\mathcal{KB}}$ be a restriction of the function $N_{C}$ associating to every $d \in N_{\D}^\mathcal{KB}$ the set $N_{C}^\mathcal{KB}(d)$ of constants  contained in $N_{C}(d)$ and in $\mathcal{KB}$. Finally, for every datatype $d \in N_{D}^\mathcal{KB}$, let $\bfk(d)$ be the set of facet expressions for $d$ occurring in $\mathcal{KB}$ and not in $N_{F}(d) \cup \{\top^{d},\bot_{d}\}$. We assume without loss of generality that the facet expressions in $\bfk(d)$ are in Conjunctive Normal Form. We define the \flqsr-formula $\phi_{\mathcal{KB}}$ expressing the consistency of $\mathcal{KB}$ as follows: {\small
\[
\phi_{\mathcal{KB}} \defAs \underset {H \in \mathcal{KB}}\bigwedge \theta(H) \wedge \bigwedge_{i=1}^{12}\xi_i   \, ,
\]}
where

\noindent$\xi_1 \defAs (\forall z)( ( \neg(z \in X_{\I}^1) \vee \neg(z \in X_{\D}^1)) \wedge (z \in X_{\D}^1 \vee  z \in X_{\I}^1))\wedge (\forall z)(z \in $

\noindent $\hfill  X_{\I}^1 \vee z \in X_{\D}^1)\wedge \neg (\forall z)\neg (z \in X_{\I}^1) \wedge \neg (\forall z)\neg (z \in X_{\D}^1)$,\\
 
 % % % % % % % % % % % % % % % % % % % % % % % % % % % % % % % % % % % %

\noindent$\xi_2 \defAs ((\forall z)( ( \neg(z \in X_{\I}^1) \vee z \in X_{\top}^1) \wedge (\neg (z \in X_{\top}^1)  \vee z \in X_{\I}^1) ) \wedge (\forall z)\neg (z \in$

\noindent $\hfill  X_{\bot})$,\\

% % % % % % % % % % % % % % % % % % % % % % % % % % % % % % % % % % % %

\noindent$\xi_3 \defAs \underset{A \in \ck}\bigwedge (\forall z)( \neg(z \in X_{A}^1) \vee z \in X_{\I}^1)$,\\

% % % % % % % % % % % % % % % % % % % % % % % % % % % % % % % % % % % %

 \noindent$\xi_4 \defAs ( \underset{d \in N_{D}^\mathcal{KB}}\bigwedge((\forall z)( \neg(z \in X_{d}^1) \vee z \in X_{\D}^1) \wedge \neg (\forall z)\neg(z \in X_{d}^1))  \wedge (\forall z)$
 
 \noindent $\hfill  (\underset{(d_i,d_j \in N_{D}^\mathcal{KB}, i < j)}\bigwedge ( ( \neg(z \in X_{d_i}^1) \vee \neg (z \in X_{d_j}^1)) \wedge ( z \in X_{d_j}^1 \vee  z \in X_{d_i}^1 ) )))$,\\
 
% % % % % % % % % % % % % % % % % % % % % % % % % % % % % % % % % % % % 

\noindent$\xi_5 \defAs \underset{d \in N_{D}^\mathcal{KB}}\bigwedge((\forall z)( ( \neg(z \in X_{d}^1) \vee z \in X_{\top_d}^1) \wedge ( \neg(z \in X_{\top_d}^1)  \vee z \in X_{d}^1  ) \wedge $

\noindent $\hfill (\forall z)\neg(z \in X_{\bot_d}^1))$,\\

% % % % % % % % % % % % % % % % % % % % % % % % % % % % % % % % % % % %

\noindent$\xi_6 \defAs   \underset {\substack{\\ f_d \in N_{F}^\mathcal{KB}(d),\\ d \in N_{D}^\mathcal{KB}}}\bigwedge (\forall z)( \neg(z \in X_{f_d}^1) \vee z \in X_{d}^1)$,\\

% % % % % % % % % % % % % % % % % % % % % % % % % % % % % % % % % % % %

\noindent$\xi_7 \defAs (\forall z_1)(\forall z_2)( ( \neg(z_1 \in X_{\I}^1) \vee \neg(z_2 \in X_{\I}^1) \vee \langle z_1,z_2 \rangle \in X_{U}^3) \wedge (    (\neg(\langle z_1,z_2 \rangle \in$

\noindent $\hfill  X_{U}^3) \vee z_1 \in X_{\I}^1) \wedge ( \neg(\langle z_1,z_2 \rangle \in X_{U}^3) \vee z_2 \in X_{\I}^1 ) ) )$,\\

% % % % % % % % % % % % % % % % % % % % % % % % % % % % % % % % % % % %

\noindent$\xi_8 \defAs  \underset {R \in \ark} \bigwedge
(\forall z_1)(\forall z_2) ( (\neg(\langle z_1,z_2\rangle \in X_R^3) \vee z_1 \in X_{\I}^1 ) \wedge ( \neg(\langle z_1,z_2\rangle \in X_R^3) \vee z_2 \in X_{\I}^1)))$,\\

% % % % % % % % % % % % % % % % % % % % % % % % % % % % % % % % % % % %

\noindent$\xi_9 \defAs \underset{T \in \crk} \bigwedge  (\forall z_1)(\forall z_2) (\neg(\langle z_1,z_2\rangle \in X_T^3) \vee z_1 \in X_{\I}^1) \wedge ( \neg(\langle z_1,z_2\rangle \in X_T^3) \vee z_2 \in X_{\D}^1)))$,\\

% % % % % % % % % % % % % % % % % % % % % % % % % % % % % % % % % % % %

\noindent$\xi_{10} \defAs \underset{a \in \ik} \bigwedge(x_a \in X_{\I}^1) \wedge \underset { \substack{ \\ d \in N_{D}^\mathcal{KB}, \\{e_d \in N_{C}^\mathcal{KB}(d)}}} \bigwedge  x_{e_d} \in X_{d}^1$,\\

% % % % % % % % % % % % % % % % % % % % % % % % % % % % % % % % % % % %

\noindent$\xi_{11} \defAs \underset {\{e_{d_1},\ldots, e_{d_n}\} \textrm{ in } \mathcal{KB}} \bigwedge (\forall z) ( ( \neg(z \in X_{\{e_{d_1},\ldots, e_{d_n}\}}^1) \vee \overset{n}  { \underset {i=1} \bigvee }(z = x_{e_{d_i}})) \wedge (  \overset{n}  { \underset {i=1} \bigwedge }(z \neq$

\noindent $\hfill  x_{e_{d_i}}  \vee  z \in X_{\{e_{d_1},\ldots, e_{d_n}\}}^1 ) ) )
 \wedge \quad \underset {\{a_{1},\ldots, a_{n}\} \textrm{ in } \mathcal{KB}} \bigwedge (\forall z)( (\neg(z \in X_{\{a_{1},\ldots, a_{n}\}}^1) \vee  $
 
 \noindent $\hfill \overset {n}{\underset {i=1} \bigvee}(z = x_{a_{i}})) \wedge (  \overset {n}{\underset {i=1} \bigwedge}(z \neq x_{a_{i}} \vee z \in X_{\{a_{1},\ldots, a_{n}\}}^1)) )$,\\
 
 % % % % % % % % % % % % % % % % % % % % % % % % % % % % % % % % % % % %

\noindent$\xi_{12} \defAs \underset { \substack{d \in N_{\D}^\mathcal{KB},\\[1.5pt] {\psi_d \in \bfk(d)}}} \bigwedge  (\forall z) ( \neg(z \in X_{\psi_d}^1) \vee z \in \zeta(X_{\psi_d}^1)) \wedge ( \neg(z \in \zeta(X_{\psi_d}^1)) \vee z \in X_{\psi_d}^1 )$

% % % % % % % % % % % % % % % % % % % % % % % % % % % % % % % % % % % %

\medskip

with $\zeta$ the transformation function from \flqsr-variables of level 1 to \flqsr-formulae recursively defined, for $d \in N_\D^\mathcal{KB}$, by
\[ {
\zeta(X_{\psi_d}^1) \defAs \begin{cases}
X_{\psi_d}^1 & \text{if } \psi_d \in N_{F}^\mathcal{KB}(d) \cup \{\top^{d},\bot_{d}\}\\
\neg \zeta(X_{\chi_d}^1) & \text{if }  \psi_d = \neg \chi_d\\
\zeta(X_{\chi_d}^1) \wedge \zeta(X_{\varphi_d}^1) & \text{if } \psi_d = \chi_d \wedge \varphi_d\\
\zeta(X_{\chi_d}^1) \vee \zeta(X_{\varphi_d}^1) & \text{if } \psi_d = \chi_d \vee \varphi_d\,.
\end{cases} }
\]
\noindent In the above formulae, the variable $X_{\I}^1$ denotes the set of individuals $\Ind$, $X_{d}^1$ a datatype $d \in N_{D}^\mathcal{KB}$, $X_{\D}^1$  a superset of the union of datatypes in $ N_{D}^\mathcal{KB}$, $X_{\top_d}^1$ and $X_{\bot_d}^1$ the constants $\top_d$ and $\bot_d$, and $X_{f_d}^1$, $X_{\psi_d}^1$ a facet $f_d$ and a facet expression $\psi_d$, for $d \in N_{D}^\mathcal{KB}$, respectively. In addition, $X_{A}^1$, $X_{R}^3$, $X_{T}^3$ denote a concept name $A$, an abstract role name $R$, and a concrete role name $T$ occurring in $\mathcal{KB}$, respectively. Finally, $X_{\{e_{d_1},\ldots,e_{d_n}\}}^1$ denotes a data range $\{e_{d_1},\ldots,e_{d_n}\}$ occurring in $\mathcal{KB}$, and $X_{\{a_{1},\ldots,a_{n}\}}^1$ a finite set $\{a_1,\ldots,a_n\}$ of nominals in $\mathcal{KB}$.

 The constraints $\xi_1-\xi_{12}$, slightly different from the constraints $\psi_1-\psi_{12}$ defined in the proof of Theorem 1 in \cite{CanLonNicSanRR2015}, are introduced to guarantee that each model of $\phi_{\KB}$ can be easily transformed in a $\shdlssx$-interpretation. 
  To prove the theorem,  we show that $\Sigma$ is the answer set for $Q$ w.r.t. $\KB$ iff $\Sigma$ is equal to $\overset{}{\underset{\M \models \phi_{\KB}}{\bigcup}}  \Sigma_{\M}'$, where $\Sigma_{\M}'$ is the collection of substitutions $\sigma$ such that $\M \models \psi_{Q}\sigma$. 
  
Preliminarly we show that if $\M$ is a $\flqsr$-interpretation such that $\M\models \phi_{\KB}$, we can construct a $\shdlss$-interpretation $\I_{\M}$ such that $\I_{\M} \models_{\D}\KB$ and, if $\I$ is a $\shdlss$-interpretation  such that $\I \models_{\D}\KB$, we can construct a  $\flqsr$-interpretation $\M_{\I}$ such that $\M_{\I}\models \phi_{\KB}$. Thus,  let $\M$ be any $\flqsr$-interpretation $\M$ such that $\M \models \phi_{\KB}$. Reasoning as in \cite{CanLonNicSanRR2015}, it is not hard to see that such $\M$ is a $\flqsr$-interpretation of the form $\M = (D_1 \cup D_2, M)$, where

\begin{itemize}
\item[-] $D_1$ and $D_2$ are disjoint nonempty sets and $\underset{d \in N_{D}^\mathcal{K}}\bigcup d^{\D} \subseteq D_2$,
\item[-]$MX_{\I}^1 \defAs D_1$, $MX_{\D}^1 \defAs D_2$, $MX_{d}^1 \defAs d^{\D}$, for every $d \in N_{D}^\mathcal{K}$, \smallskip
\item [-] $MX_{f_d}^1 \defAs f_d^{\D}$, for every $f_d \in N_{F}^\mathcal{K}(d)$, with $d \in N_{D}^\mathcal{K}$.
\end{itemize}

\noindent

Exploiting the fact that $\M$ satisfies the constrains $\xi_1-\xi_{12}$, it is then possible to define a $\shdlssx$-interpretation $\IM=(\Delta^I,\Delta_\D,\cdot^I)$ , by putting 
\begin{itemize}
\item $\Delta^{\I} \defAs MX_{\I}^1$, 
\item $\Delta_{\D} \defAs MX_{\D}^1$, 
\item $A^\I \defAs MX_{A}^1$, for every concept name $A \in \ck$, 
\item $S^\I \defAs \{ \langle u_1, u_2 \rangle : u_1 \in MX^1_I, u_2 \in MX^1_I, \langle u_1,u_2 \rangle \in MX^3_S \}$, for every abstract role name $S \in \ark$, 
\item $T^\I \defAs \{ \langle u_1, u_2 \rangle : u_1 \in MX^1_I, u_2 \in MX^1_D, \langle u_1,u_2 \rangle \in MX^3_T \}$, for every concrete role name $T \in \crk$,
\item $a^\I \defAs Mx_{a}$, for every individual $a \in \ik$,
\item $e_{d}^D \defAs Mx_{e_{d}}$, for every constant $e_{d} \in N_{C}^{\KB}(d)$ with $d \in N_{D}^{\KB}$.
\end{itemize}

Since $\M \models  \underset{H \in \KB}{\theta(H)} \wedge \overset{12}{\underset{i=1}{\bigwedge}} \xi_i$, and, as it can easily checked, $\IM \models_{\D} H$,  iff $\M \models \theta(H)$, for every statement $H \in \KB$, we plainly have that $\IM \models_{\D} \KB$. Conversely, let $\I = (\Delta^{\I}, \Delta_{\D}, \cdot^{\I})$ be a $\shdlssx$-interpretation  such that $\I \models_{\D} \KB$. We show how to construct, out of the datatype map $\D$ and the $\shdlssx$-interpretation $\I$, a \flqsr-interpretation $\mathbfcal{M}_{\I,\D} = (D_{\I,\D}, M_{\I,\D})$ which satisfies $\phi_{\KB}$. Let us put $D_{\I,\D} \defAs \Delta^{\I} \cup \Delta_{\D}$ and define $M_{\I,\D}$ by putting $M_{\I,\D} X_{\I}^1 \defAs \Delta^{\I}$, $M_{\I,\D} X_{\D}^1 \defAs \Delta_{\D}$,  $M_{\I,\D} X_{U}^3 \defAs U^{\I}$, $M_{\I,\D} X_{dr}^1 \defAs dr^{\D}$, for every variable $X_{dr}^1$ in $\phi_\KB$ denoting a data range $dr$ occurring in $\KB$, $M_{\I,\D} X_{A}^1 \defAs A^{\I}$, for every $X_{A}^1$ in $\phi_\KB$ denoting a concept name in $\KB$, and $M_{\I,\D} X_{S}^3 \defAs S^{\I}$, for every $X_{S}^3$ in $\phi_\KB$ denoting an abstract role name in $\KB$. Variables $X_{T}^3$, denoting concrete role names, and variables $x_a, x_{e_d}$, denoting individuals and datatype constants, respectively, are interpreted in a similar way. From the definitions of $\D$ and $\I$, it follows easily that  $\mathbfcal{M}_{\I,\D}$ satisfies the formulae $\xi_1$-$\xi_{12}$ and $\theta(H)$, for every statement $H \in \KB$, and, therefore, that $\mathbfcal{M}_{\I,\D}$ is a model for $\phi_{\KB}$.

Now we prove the first part of the theorem. Let us assume that $\Sigma$ is the answer set for $Q$ w.r.t. $\KB$. We have to show that $\Sigma$ is equal to $ \Sigma' = \underset{\M \models \phi_{\KB}}{\bigcup} \Sigma'_{\M}$, where $\Sigma'_{\M}$ is the collection of all the substitutions $\sigma'$ such that $\M \models \psi_{Q}\sigma'$.
 
By contradiction, let us assume that there exists a $\sigma \in \Sigma$ such that $\sigma \notin \Sigma'$, namely $\M \not\models \psi_{Q}\sigma$, for every $\flqsr$-interpretation $\M$ with $\M \models \phi_{\KB}$. Since $\sigma \in \Sigma$ there is a $\shdlssx$-interpretation $\I$ such that $\I \models_{\D} \KB$ and $\I \models_{\D} Q\sigma$. Then, by the construction above, we can define a $\flqsr$-interpretation $\M_{\I}$ such that $\M_{\I} \models \phi_{\KB}$ and $\M_{\I}\models \psi_Q\theta{\sigma}$. Absurd. 

Conversely, let $\sigma' \in \Sigma'$ and assume by contradiction that $\sigma' \notin \Sigma$. 
Then, for all $\shdlssx$-interpretations such that $\I \models_D \KB$, it holds that $\I \not\models_D Q\sigma'$. 
Since $\sigma'\in \Sigma'$, there is a $\flqsr$-interpretation $\M$ such that $\M \models \phi_{\KB}$ and $\M \models \psi\sigma'$. Then, by the construction above, we can define a $\shdlss$-interpretation $\I_{\M}$ such that  $\I_{\M} \models_D \KB$ and $\I_{\M} \models_D Q\sigma'$. Absurd.  

% 
%Details of the construction of $\theta$ and of $\xi_1$--$\xi_{12}$ can be found in \cite{ExtendedVersionICTCS2016}.}

%
%
% can be mapped into $\flqsr$-formulae $\phi_{\KB}$ and $\psi_Q$ 
%
%
%in $\flqsr$-formulae in conjunctive normal form (CNF), and the answer set $\Sigma$ for $Q$ w.r.t. $\KB$ in a set $\Sigma'$ of substitutions in the context of the fragment $\flqsr$. More specifically, by means of $\theta$ we can construct the $\flqsr$-formulae  
%\[
%\varphi_{\KB} \defAs \bigwedge_{H \in \KB} \theta(H) \wedge \bigwedge_{i=1}^{12} \xi_i, \qquad \psi_Q \defAs \theta(Q)
%\]
%%$\varphi_{\KB} \defAs \bigwedge_{H \in \KB} \theta(H) \wedge \bigwedge_{i=1}^{12} \xi_i$ and $\psi_Q \defAs \theta(Q)$, 
%and the set $\Sigma' \defAs \{\theta(\sigma) : \sigma \in \Sigma\}$. 
%%
%Then it can be shown that $\Sigma$ is the answer set for $Q$ w.r.t. $\KB$ iff $\Sigma'$ is equal to $\bigcup_{\M \models \varphi_{\KB}}  \Sigma'_{\M}$, where $\Sigma'_{\M}$ is the collection of substitutions $\sigma'$ such that $\M \models \psi_{Q}\sigma'$. 
\end{proof}

\section{A tableau-based procedure} \label{tableaucon}
In this section, we illustrate a \ke\space based procedure that, given a $\flqsr$-formula $\phi_\KB$ corresponding to a $\shdlssx$-knowledge base and a $\flqsr$-formula $\psi_Q$ corresponding to a $\shdlssx$-conjunctive query $Q$, yields all the substitutions $\sigma = \{x_1/y_1,\ldots,x_n/y_n\}$, with $\{x_1,\ldots,x_n\} =  \varz(\psi_Q)$ and $\{y_1,\ldots,y_n\}\subseteq \varz(\phi_\KB)$, belonging to the answer set $\Sigma'$ of $\psi_Q$ w.r.t.\ $\phi_\KB$. 

Let $\overline{\phi}_\KB$ be the formula obtained from $\phi_\KB$ by:
 \begin{itemize}[topsep=0.1cm, itemsep=0.cm]
 \item[-]  moving universal quantifiers in $\phi_\KB$ as inwards as possible  according to the  rule $(\forall z)(A(z) \wedge B(z)) \longleftrightarrow ((\forall z) A(z) \wedge (\forall z) B(z))$,
 \item[-] renaming universally quantified variables so as to make them pairwise distinct.
 \end{itemize}

Let $F_1, \ldots, F_k$ be the conjuncts of  $\overline{\phi}_\KB$ that are $\flqsr$-quantifier-free atomic formulae and $S_1, \ldots, S_m$ the conjuncts of  $\overline{\phi}_\KB$ that are $\flqsr$-purely universal formulae. For every $S_i = (\forall z_1^i) \ldots (\forall z_{n_i}^i) \chi_i$, $i=1,\ldots,m$, we put\\[.4cm]
\centerline{$Exp(S_i) \defAs \underset{ \{x_{a_1}, \ldots, x_{a_{n_i}}\} \subseteq \varz(\overline{\phi}_{\KB})}{\bigwedge} S_i \{z_1^i / x_{a_1}, \ldots, z^i_{n_i} / x_{a_{n_i}} \}$.}\\[.3cm]
%, where $\ik$ is the set of individual names in $\KB$.
Let $\Phi_\KB \defAs \{ F_j : i=1,\ldots,k \} \cup \overset{m}{ \underset{i=1}{\bigcup}} Exp(S_i)$.

To prepare for the \ke\space based procedure to be described next, we introduce some  useful  notions and notations (see \cite{dagostino1999} for a detailed overview of \ke, an optimized variant of semantic tableaux).
%It is convenient to give some notions useful for the definition of a \ke\space  like procedure introduced next to  construct the models of $\Phi_\KB$. 
%In what follows we give the definition of a \ke\space for a  collection of clauses of $\flqsr$-quantifier free atomic formulae of level $0$.

Let $\Phi = \{ C_1,\ldots, C_p\}$ be a collection of disjunctions of $\flqsr$-quantifier-free atomic formulae of level $0$ of the types: $x =y$, $x \in X^1$, $\langle x,y\rangle \in X^3$. $\mathcal{T}$ is a \textit{\ke}\space for $\Phi$ if there exists a finite sequence $\mathcal{T}_1, \ldots, \mathcal{T}_t$ such that (i) $\mathcal{T}_1$ is a one-branch tree consisting of the sequence $C_1,\ldots, C_p$, (ii) $\mathcal{T}_t = \mathcal{T}$, and (iii) for each $i<t$, $\mathcal{T}_{i+1}$ is obtained from $\mathcal{T}_i$ by an application of one of the  rules in Fig \ref{exprule}. The set of formulae $\seq = \{ \overline{\beta}_1,\ldots,\overline{\beta}_n\} \setminus \{\overline{\beta}_i\}$ occurring as premise in the E-rule contains the  complements of all the components of the formula $\beta$ with the exception of the component $\beta_i$.
 
%
%
%\vspace{-.2cm}
\begin{figure}
{
\begin{center}
\begin{minipage}[h]{5cm}
$\infer[\textbf{E-Rule}]
{\beta_i}{\beta_1 \vee \ldots \vee \beta_n & \quad \seq}$\\[.1cm]
{ where $\seq \defAs \{ \overline{\beta}_1,...,\overline{\beta}_n\} \setminus \{\overline{\beta}_i\}$,}\\[-.1cm] { for $i=1,...,n$}
\end{minipage}~~~~~~~~~~~
\begin{minipage}[h]{2.5cm}
\vspace{0.1cm}
$\infer[\textbf{PB-Rule}]
{A~~|~~\overline{A}}{}$\\[.1cm]
{ with $A$ a literal}
\end{minipage}
\end{center}
\vspace{-.2cm}
}
%\begin{center} 
%\begin{minipage}[]{5cm}
%\raggedright
%\vspace{0.4cm}
%\[
%  \frac{\begin{array}{@{}c@{}}
%    \beta_1 \vee \ldots \vee \beta_n  \\
%    \seq
%  \end{array}}{
%    \beta_i } \textbf{E-Rule}
%\]
%\centering with $\seq = \{ \overline{\beta}_1,...,\overline{\beta}_n\} \setminus \{\overline{\beta}_i\}$\\$i=1,...,n$
%\end{minipage}
%\raggedright
%\begin{minipage}[]{5cm}
%\[
%  \frac{\begin{array}{@{}c@{}}
%   \\
%  \end{array}}{
%  \hspace{0.5cm} A \hspace{0.3cm} \mid  \hspace{0.3cm}\overline{A}\hspace{0.5cm} } \textbf{PB-Rule}
%\]
%\centering with $A$ literal\\
%\end{minipage}
%\end{center}

%\begin{center} 
%\begin{minipage}[]{5cm}
%\raggedleft
%\[
%  \frac{\begin{array}{@{}c@{}}
%   \hspace{0.2cm}X \\
%   \neg X
%  \end{array}}{
%   \bot } \textbf{Clash-Rule}
%\]
%\end{minipage}
%\raggedright
%\begin{minipage}[]{5cm}
%\vspace{0.3cm}
%\[
%  \frac{\begin{array}{@{}c@{}}
%   \neg X \\
%   \hspace{0.2cm} X
%  \end{array}}{
%   \bot } \textbf{Clash-Rule}
%\]
%\end{minipage} \vspace{0.3cm} 
\caption{\label{exprule}Expansion rules for the \ke.}
% \end{center}
\normalsize
\end{figure}

Let $\mathcal{T}$ be a \ke. A branch $\theta$ of $\mathcal{T}$  is \textit{closed} if it contains both $A$ and $\neg A$, for some formula $A$. Otherwise, the branch is \textit{open}. A formula $\beta_1 \vee \ldots \vee \beta_n$ is \textit{fulfilled} in a branch $\vartheta$, if $\beta_i$ is in $\theta$, for some $i=1,\ldots,n$. A branch $\vartheta$ is \textit{complete} if every formula $\beta_1 \vee \ldots \vee \beta_n$ occurring in $\vartheta$ is fulfilled. A \ke\space is \textit{complete} if all its  branches are complete.

Next we introduce the procedure Saturate-KB that takes as input the set $\Phi_{\KB}$ constructed from a $\flqsr$-formula $\phi_\KB$ representing a $\shdlssx$-knowledge base $\KB$ as shown above, and yields a complete \ke\space $\mathcal{T}_{\KB}$ for $\Phi_\KB$.

{
\begin{procedure}\label{proced}
Saturate-KB($\Phi_\KB$)
%$T_{\KB}$ = $\Phi_\KB$;
%\vspace{-.2cm}
\begin{enumerate} 
\item $\T_{\KB}$ := $\Phi_\KB$;
\item Select an open branch $\vartheta$ of $\T_{\KB}$ that is not yet complete.
\begin{enumerate}
\item Select a formula $\beta_1 \vee \ldots \vee \beta_n$ on $\vartheta$ that is not fulfilled.
\item If $\seqnj$ is in $\vartheta$, for some $j \in \{1,\ldots,n\}$, apply the E-Rule to $\beta_1 \vee \ldots \vee \beta_n$ and $\seqnj$ on $\vartheta$ and go to step 2.
\item If $\seqnj$ is not in $\vartheta$, for every $j=1,\ldots,n$, let $B^{\overline{\beta}}$ be the collection of formulae $\overline{\beta}_1,\ldots,\overline{\beta}_n$ present in $\vartheta$
%subset of $\{\overline{\beta}_1,\ldots,\overline{\beta}_n\}$ that  is included in $\vartheta$ 
and let $\overline{\beta}_h$ be the lowest index formula such that $\overline{\beta}_h \in \{\{ \overline{\beta}_1,\ldots,\overline{\beta}_n  \} \setminus B^{\overline{\beta}}\}$, then apply the PB-rule to $\overline{\beta}_h$ on $\vartheta$, and go to step 2.
\end{enumerate}
\item Return $\T_{\KB}$.
\end{enumerate}
\end{procedure}
}
\normalsize

Soundness of Procedure \ref{proced} can be easily proved in a standard way and its completeness can be shown much along the lines of Proposition 36 in \cite{dagostino1999}. 
Concerning termination of Procedure \ref{proced}, our proof is based on the following two facts. %The number of non-fulfilled formulae on the one-branch initial \ke\space is obviously finite, since $|\Phi_{\KB}|$ is finite. 
The rules in Fig. \ref{exprule} are applied only to non-fulfilled formulae on open branches and tend to reduce the number of non-fulfilled formulae occurring on the considered branch. In particular, when the E-Rule is applied on a branch $\vartheta$, the number of non-fulfilled formulae on $\vartheta$ decreases. In case of application of the PB-Rule on a formula $\beta = \beta_1 \vee \ldots \vee \beta_n$ on a branch, the rule generates two branches. In one of them the number of non-fulfilled formulae decreases (because $\beta$ becomes fulfilled).  In  the other one the number of non-fulfilled formulae stays constant but the subset $B^{\overline{\beta}}$ of $\{\overline{\beta}_1,\ldots,\overline{\beta}_n\}$
occurring on the branch gains a new element. Once $|B^{\overline{\beta}}|$ gets equal to $n-1$, namely after at most $n-1$ applications of the PB-rule, the E-rule is applied and the formula $\beta = \beta_1 \vee \ldots \vee \beta_n$ becomes fulfilled, thus decrementing the number of non-fulfilled formulae on the branch. Since the number of non-fulfilled formulae on each open branch gets equal to zero after a finite number of steps and the rules of Fig. \ref{exprule} can be applied only to non-fulfilled formulae on open branches,  the procedure terminates.

By the completeness of Procedure \ref{proced}, each branch $\vartheta$ of $\T_{\KB}$ induces a $\flqsr$-interpretation $\M_{\vartheta}$ such that $\M_{\vartheta} \models \Phi_{\KB}$. 
We define $\M_\vartheta = (D_\vartheta, M_\vartheta)$ as follows. We put

\begin{itemize}[itemsep=0.2cm]
\item [-] $D_\vartheta\defAs\{ x \in \mathcal{V}_0 : x\mbox{ occurs in }\vartheta\}$;
\item [-]  $M_\vartheta x \defAs x$,  for every $x \in  D_\vartheta$;
\item [-] $M_\vartheta X_{C}^1= \{ x : x \in X_{C}^1 \mbox{ is in } \vartheta \}$, for every $X_{C}^1 \in \mathcal{V}_1$ occurring $\vartheta$;
\item [-]$M_\vartheta X_{R}^3= \{ \langle x,y \rangle :  \langle x,y \rangle \in X_{R}^3 \mbox{ is in } \vartheta \}$, for every $X_{R}^3 \in \mathcal{V}_3$ occurring in $\vartheta$.
\end{itemize}

It is easy to check that $\M_\vartheta\models \overline{\phi}_{\KB}$ and thus, plainly, that $\M_\vartheta\models \phi_{\KB}$.
Next, we provide some complexity results.
Let $r$ be the maximum number of universal quantifiers in $S_i$, and $k \defAs |\varz(\overline{\phi}_{\KB})|$. Then, each $S_i$ generates $k ^r$ expansions. Since the knowledge base contains $m$ such formulae, the number of disjunctions in the initial branch of the \ke\space is $m \cdot k^r$. Next, let $\ell$ be the maximum number of literals in $S_i$, for $i=1,\ldots,m$. Then, the maximum depth of the \ke, namely the maximum size of the models of $\Phi_{\KB}$ constructed as illustrated above, is $\mathcal{O}(  \ell  m  k^r)$ and the number of leaves of the tableau, that is the number of such models of $\Phi_{\KB}$, is $O(2^{\ell  m  k^r})$.

We now describe a procedure that, given a \ke\space constructed by Procedure \ref{proced} and a $\flqsr$-formula $\psi_Q$ representing a $\shdlssx$-conjunctive query $Q$, yields all the substitutions $\sigma'$ in the answer set $\Sigma'$ of $\psi_{Q}$ w.r.t.\ $\phi_\KB$. By the soundness of Procedure \ref{proced}, we can limit ourselves to consider only the models $\M_\vartheta$ of $\phi_\KB$ induced by each open branch $\vartheta$ of $\T_{\KB}$. For every open and complete branch $\vartheta$ of $\T_{\KB}$, we construct a decision tree $\DT_{\vartheta}$ such that every maximal branch of $\DT_{\vartheta}$ defines a substitution $\sigma'$ such that  $\M_\vartheta \models \psi_Q\sigma'$. 

Let $d$ be the number of literals in $\psi_Q$. $\DT_\vartheta$ is a finite labelled tree of depth $d+1$ whose labelling satisfies the following conditions, for $i=0,\ldots,d$:
\begin{itemize}[itemsep=0.2cm]
\item[(i)]  every node of $\DT_\vartheta$ at level $i$ is labelled with $(\sigma_i, \psi_Q\sigma_i)$, and, in particular, the root is labelled with
$(\sigma'_0, \psi_Q\sigma'_0)$, where $\sigma'_0$ is the empty substitution; 

\item[(ii)] if a node at level $i$ is labelled with $(\sigma'_i, \psi_Q\sigma'_i)$, then its $s$-successors, with $s >0$, are labelled with \\[0.2cm]
\centerline{
$\big(\sigma'_i\varrho^{q_i+1}_1, \psi_Q(\sigma'_i\varrho^{q_i+1}_1)\big),\ldots,\big(\sigma'_i\varrho^{q_i+1}_s, \psi_Q(\sigma'_i\varrho^{q_i+1}_s)\big)$, 
}\\[0.2cm]
where $q_{i+1}$ is the $(i+1)$-st conjunct of $\psi_Q\sigma'_i$ and $\mathcal{S}_{q_{i+1}}=\{\varrho^{q_i+1}_1, \ldots, \varrho^{q_i+1}_s  \}$ is the collection of the substitutions $\varrho = \{x_1/y_1,\ldots,x_j/ y_j\}$ with $\{x_1,\ldots,x_j\} = \varz(q_{i+1})$ such that $p=q_{i+1}\varrho$, for some  literal $p$ on $\vartheta$. If $s = 0$, the node labelled with $(\sigma'_i, \psi_Q\sigma'_i)$ is a leaf node and, if $i = d$,  $\sigma'_i$ is added to $\Sigma'$.  

\end{itemize}

%respectively 

Let $\delta(\T_{\KB})$ and $\lambda(\T_{\KB})$ be, respectively, the maximum depth of $\T_{\KB}$ and the number of leaves of $\T_{\KB}$ computed above. Plainly, $\delta(\T_{\KB}) = \mathcal{O}(\ell  m  k^r)$  and $\lambda(\T_{\KB})= \mathcal{O}(2^{\ell  m  k^r})$. It is easy to verify that $s=2^{k}$ is the maximum branching of $\DT_\vartheta$. Since $\DT_\vartheta$ is a $s$-ary  tree of depth $d+1$, where $d$ is the number of literals in $\psi_Q$, and the $s$-successors of a node are computed in $\mathcal{O}(\delta(\T_{\KB}))$ time, the number of leaves in $\DT_\vartheta$ is $\mathcal{O}(s^{(d+1)})=\mathcal{O}(2^{k (d+1)})$ and they are computed  in $\mathcal{O}(2^{k  (d+1)}  \delta(\T_{\KB}))$ time. Finally, since we have $\lambda(\T_{\KB})$ of such decision trees, the answer set of $\psi_{Q}$ w.r.t.\ $\phi_\KB$ is computed in time\\

\centerline{$\mathcal{O}(2^{k  (d+1)}  \delta(\T_{\KB})  \lambda(\T_{\KB}))=$ $\mathcal{O}(  2^{k (d+1)} \cdot \ell  m  k^r \cdot 2^{\ell  m  k^r})= \mathcal{O}(\ell  m  k^r  2^{k (d+1) +\ell  m  k^r} )$ .} \medskip

 Since the size of $\phi_\KB$ and of $\psi_{Q}$ are polynomially related to those of $\KB$ and of $Q$, respectively (see \cite{ExtendedVersionICTCS2016} for details on the reduction), the construction of the answer set of $Q$ with respect to $\KB$ can be done in double-exponential time. In case $\KB$ contains no role chain axioms and qualified cardinality restrictions, the complexity of our CQA problem is in EXPTIME, since the maximum number of universal quantifiers in $\phi_{\KB}$, namely $r$, is a constant (in particular $r = 3$).  
We remark that such result is comparable with the complexity of the CQA problem for a large family of description logics such as $\mathcal{SHIQ}$ \cite{Ortiz:2011:QAH:2283516.2283571}. In particular, the CQA problem for the very expressive description logic $\mathcal{SROIQ}$ turns out to be 2-NEXPTIME-complete.  

\section{Conclusions}
We have introduced the description logic $\dlssx$ ($\shdlssx$, for short) that extends the logic $\dlss$ with Boolean operations on concrete roles and on the product of concepts. 
%and \red{we proved its decidability} by resorting to the decision procedure for the satisfiability problem of  a four-level stratified syllogistic called $\flqsr$. 
We addressed the problem of Conjunctive Query Answering for the description logic $\shdlssx$ by formalizing $\shdlssx$-knowledge bases and $\shdlssx$-conjunctive queries in terms of formulae of $\flqsr$. Such formalization seems to be  promising for implementation purposes. 

In our approach, we first constructed a \ke\space $\T_{\KB}$ for $\phi_{\KB}$, a $\flqsr$-formalization of a given $\shdlssx$-knowledge base $\KB$, whose branches induce the models of $\phi_{\KB}$.  
%We chose the  \ke\space system because it prevents the construction of redundant branches and thus it is more suitable for implementation. 
Then we computed the answer set of a  $\flqsr$-formula $\psi_{Q}$, representing a $\shdlssx$-conjunctive query $Q$, with respect to  $\phi_{\KB}$  by means of a forest of decision trees based on the branches of $\T_{\KB}$ and gave some complexity results.  
%We also showed that the complexity of construction for the models of $\phi_{\KB}$ is 2-EXPTIME in the size of $\phi_{\KB}$, and that the computation of the answer set of $\psi_Q$ with respect to $\phi_{\KB}$ is 2-EXPTIME in the size of $\phi_{\KB}$ and $\psi_{Q}$.

We plan to generalize our procedure with a data-type checker in order to extend reasoning with data-types, and also to extend $\flqsr$ with data-type groups. We also intend to improve the efficiency of the knowledge base saturation algorithm and query answering algorithm, and to extend the expressiveness of the queries. Finally, we intend to study a parallel model of the  procedure described and to provide an implementation of it.
\bibliographystyle{ieeetran}
\bibliography{biblio} 

% Generated by IEEEtranS.bst, version: 1.14 (2015/08/26)
\begin{thebibliography}{10}
\providecommand{\url}[1]{#1}
\csname url@samestyle\endcsname
\providecommand{\newblock}{\relax}
\providecommand{\bibinfo}[2]{#2}
\providecommand{\BIBentrySTDinterwordspacing}{\spaceskip=0pt\relax}
\providecommand{\BIBentryALTinterwordstretchfactor}{4}
\providecommand{\BIBentryALTinterwordspacing}{\spaceskip=\fontdimen2\font plus
\BIBentryALTinterwordstretchfactor\fontdimen3\font minus
  \fontdimen4\font\relax}
\providecommand{\BIBforeignlanguage}[2]{{%
\expandafter\ifx\csname l@#1\endcsname\relax
\typeout{** WARNING: IEEEtranS.bst: No hyphenation pattern has been}%
\typeout{** loaded for the language `#1'. Using the pattern for}%
\typeout{** the default language instead.}%
\else
\language=\csname l@#1\endcsname
\fi
#2}}
\providecommand{\BIBdecl}{\relax}
\BIBdecl

\bibitem{calvanese2007answering}
D.~Calvanese, T.~Eiter, and M.~Ortiz, ``Answering regular path queries in
  expressive description logics: An automata-theoretic approach,'' in
  \emph{AAAI}, vol.~7, 2007, pp. 391--396.

\bibitem{Calvanese2013335}
D.~Calvanese, G.~D. Giacomo, D.~Lembo, M.~Lenzerini, and R.~Rosati, ``Data
  complexity of query answering in description logics,'' \emph{Artificial
  Intelligence}, vol. 195, pp. 335 -- 360, 2013.

\bibitem{Calvanese:1998:DQC:275487.275504}
D.~Calvanese, G.~D. Giacomo, and M.~Lenzerini, ``On the decidability of query
  containment under constraints,'' in \emph{Proc. of the Seventeenth ACM
  SIGACT-SIGMOD-SIGART Symposium on Principles of Database Systems}, ser. PODS
  '98.\hskip 1em plus 0.5em minus 0.4em\relax New York, NY, USA: ACM, 1998, pp.
  149--158.

\bibitem{CanNic2013}
D.~Cantone and M.~Nicolosi-Asmundo, ``On the satisfiability problem for a
  4-level quantified syllogistic and some applications to modal logic,''
  \emph{Fundamenta Informaticae}, vol. 124, no.~4, pp. 427--448, 2013.

\bibitem{ExtendedVersionICTCS2016}
D.~Cantone, M.~Nicolosi-Asmundo, and D.~Santamaria, ``{Conjunctive Query
  Answering via a Fragment of Set Theory (Extended Version)},'' 2016, available
  at https://archive.org/details/CanNicSan16\_Extended.

\bibitem{CanLon2014}
D.~Cantone and C.~Longo, ``A decidable two-sorted quantified fragment of set
  theory with ordered pairs and some undecidable extensions,'' \emph{Theor.
  Comput. Sci.}, vol. 560, pp. 307--325, 2014.

\bibitem{CanLonNicSanRR2015}
D.~Cantone, C.~Longo, M.~N. Asmundo, and D.~F. Santamaria, ``Web ontology
  representation and reasoning via fragments of set theory,'' in \emph{Web
  Reasoning and Rule Systems - 9th International Conference, {RR} 2015, Berlin,
  Germany, August 4-5, 2015, Proceedings}, 2015, pp. 61--76.

\bibitem{CanLonNic2010}
D.~Cantone, C.~Longo, and M.~Nicolosi~Asmundo, ``A decision procedure for a
  two-sorted extension of multi-level syllogistic with the {C}artesian product
  and some map constructs,'' in \emph{Proc.\ of the 25th Italian Conference on
  Computational Logic (CILC 2010), Rende, Italy, July 7-9, 2010}, W.~Faber and
  N.~Leone, Eds., vol. 598.\hskip 1em plus 0.5em minus 0.4em\relax CEUR
  Workshop Proceedings, ISSN 1613-0073, June 2010, pp. 1--18 (paper 11).

\bibitem{CanLonNic11}
D.~Cantone, C.~Longo, and M.~{Nicolosi-Asmundo}, ``A decidable quantified
  fragment of set theory involving ordered pairs with applications to
  description logics,'' in \emph{Proc. Computer Science Logic, 20th Annual
  Conf. of the EACSL, {CSL} 2011, September 12-15, 2011, Bergen, Norway}, 2011,
  pp. 129--143.

\bibitem{CanLonPis2010}
D.~Cantone, C.~Longo, and A.~Pisasale, ``Comparing description logics with
  multi-level syllogistics: the description logic
  $\mathcal{DL}\langle\mathsf{MLSS}_{2,m}^{\times}\rangle$$\mathcal{DL}\langle\mathsf{MLSS}_{2,m}^{\times}\rangle$,''
  in \emph{6th Workshop on Semantic Web Applications and Perspectives
  (Bressanone, Italy, Sep. 21-22, 2010)}, P.~Traverso, Ed., 2010, pp. 1--13.

\bibitem{dagostino1999}
M.~D'Agostino, ``Tableau methods for classical propositional logic,'' in
  \emph{Handbook of Tableau Methods}, M.~D'Agostino, D.~M. Gabbay,
  R.~H\"{a}hnle, and J.~Posegga, Eds.\hskip 1em plus 0.5em minus 0.4em\relax
  Springer, 1999, pp. 45--123.

\bibitem{DeJo90}
N.~Dershowitz and J.-P. Jouannaud, ``Rewrite systems,'' in \emph{Handbook of
  Theoretical Computer Science (Vol. B)}, J.~van Leeuwen, Ed.\hskip 1em plus
  0.5em minus 0.4em\relax Cambridge, MA, USA: MIT Press, 1990, pp. 243--320.

\bibitem{GlHS07a}
B.~Glimm, I.~Horrocks, and U.~Sattler, ``Conjunctive query entailment for
  shoq,'' in \emph{Proceedings of the 2007 Description Logic Workshop (DL
  2007)}, vol. 250, 2007, pp. 1--11.

\bibitem{GliHoLuSa-JAIR08}
B.~Glimm, C.~Lutz, I.~Horrocks, and U.~Sattler, ``Answering conjunctive queries
  in the $\mathcal{SHIQ}$ description logic,'' \emph{Journal of Artificial
  Intelligence Research}, vol.~31, pp. 150--197, 2008.

\bibitem{HorrSattTob-CADE-2000}
I.~{Horrocks}, U.~{Sattler}, and S.~{Tobies}, ``Reasoning with individuals for
  the description logic shiq,'' in \emph{Proceedings of the 17th International
  Conference on Automated Deduction {(CADE-17)}}, ser. Lecture Notes in
  Computer Science, D.~{MacAllester}, Ed., no. 1831.\hskip 1em plus 0.5em minus
  0.4em\relax Germany: Springer Verlag, 2000, pp. 482--496.

\bibitem{HorrTess-aaai-2000}
I.~{Horrocks} and S.~{Tessaris}, ``A conjunctive query language for description
  logic aboxes,'' in \emph{In Proc. of the 17th Nat. Conf. on Artificial
  Intelligence {(AAAI-2000)}}, 2000, pp. 399--404.

\bibitem{DBLP:conf/ijcai/HustadtMS05}
U.~Hustadt, B.~Motik, and U.~Sattler, ``Data complexity of reasoning in very
  expressive description logics,'' in \emph{IJCAI-05, Proc. of the Nineteenth
  International Joint Conference on Artificial Intelligence, Edinburgh,
  Scotland, UK, July 30-August 5, 2005}, 2005, pp. 466--471.

\bibitem{DBLP:conf/cade/Lutz08}
C.~Lutz, ``The complexity of conjunctive query answering in expressive
  description logics,'' in \emph{Automated Reasoning, 4th International Joint
  Conference, {IJCAR} 2008, Sydney, Australia, August 12-15, 2008,
  Proceedings}, 2008, pp. 179--193.

\bibitem{Motik2008}
B.~Motik and I.~Horrocks, ``{OWL} datatypes: Design and implementation,'' in
  \emph{Proc. of the 7th Int. Semantic Web Conference (ISWC 2008)}, ser. LNCS,
  vol. 5318.\hskip 1em plus 0.5em minus 0.4em\relax Springer, October 26--30
  2008, pp. 307--322.

\bibitem{j.websem63}
B.~Motik, U.~Sattler, and R.~Studer, ``Query answering for owl-dl with rules,''
  \emph{Web Semantics: Science, Services and Agents on the World Wide Web},
  vol.~3, no.~1, pp. 41--60, 2005.

\bibitem{Ortiz:Calvanese:et-al:06a}
M.~Ortiz, D.~Calvanese, and T.~Eiter, ``Characterizing data complexity for
  conjunctive query answering in expressive description logics,'' in
  \emph{Proc.\ of the 21st Nat.\ Conf.\ on Artificial Intelligence
  (AAAI~2006)}, 2006, pp. 1--6.

\bibitem{Ortiz:2011:QAH:2283516.2283571}
M.~Ortiz, R.~Sebastian, and M.~\v{S}imkus, ``Query answering in the horn
  fragments of the description logics shoiq and sroiq,'' in \emph{Proc. of the
  Twenty-Second International Joint Conference on Artificial Intelligence -
  Volume Volume Two}, ser. IJCAI'11.\hskip 1em plus 0.5em minus 0.4em\relax
  AAAI Press, 2011, pp. 1039--1044.

\bibitem{Rosa07c}
R.~Rosati, ``On conjunctive query answering in {EL},'' in \emph{Proceedings of
  the 2007 International Workshop on Description Logic (DL~2007)}.\hskip 1em
  plus 0.5em minus 0.4em\relax CEUR Electronic Workshop Proceedings, 2007, pp.
  1--8.

\bibitem{owl2spec}
{World Wide Web Consortium}, ``{OWL} 2 {W}eb ontology language structural
  specification and functional-style syntax (second edition).''

\end{thebibliography}
\end{document}